\documentclass[a4paper,10pt]{article}
\usepackage[top=3cm,right=2.2cm,bottom=2.5cm,left=2.2cm]{geometry} 
\usepackage{setspace}
\usepackage{hyperref}
\usepackage{amsmath}
\usepackage{amsthm}
\usepackage{amssymb}
\usepackage{amsfonts}
\usepackage{eufrak}

\usepackage{mathrsfs}
\usepackage{mathtools}
\usepackage{graphicx} %
\usepackage{cite}
\usepackage{verbatim}
\usepackage{hyperref}%
\usepackage{xcolor}
\usepackage{tikz}
\usetikzlibrary{calc,matrix,fit}
\usepackage{colortbl}
\usetikzlibrary{shapes,decorations}
\usepackage[textsize=scriptsize]{todonotes}
\usepackage[]{graphicx}
\usepackage{caption}
\usepackage{subcaption}
\usepackage{wrapfig}
\usepackage{authblk}
\usepackage{nicematrix}

\DeclareMathOperator\arctanh{arctanh}

\makeatother

\NewDocumentCommand{\Frame}{}{\Block[draw, fill=yellow!30, line-width=2pt]{1-1}{}}

\renewcommand{\Re}{\operatorname{Re}}
\newcommand{ \ud }{\,\mathrm{d}}

\newcommand{\RSB}{\boldsymbol{\mathcal{F}}}

\newcommand{\Ptl}{\mathcal{Z}}

\newcommand{\Lie}{\mathcal{L}}

\DeclareMathOperator*{\Z}{\mathcal Z}

\newcommand{\tscomo}[1]{}

\title{Interacting Kerr-Newman electromagnetic fields}
\author[1]{Sajad Aghapour\thanks{sajad.aghapour@aei.mpg.de}}
\affil[1]{Max Planck Institute for Gravitational Physics (Albert Einstein
Institute), D-14476 Potsdam, Germany}
\author[2]{Lars Andersson\thanks{lars.andersson@bimsa.cn}}
\affil[2]{Beijing Institute of Mathematical Sciences and Applications, Beijing 101408, China}
\author[3]{Kjell Rosquist\thanks{kr@fysik.su.se}}
\affil[3]{Department of Physics, Stockholm University, Albanova University Center, SE-106 91 Stockholm, Sweden}
\author[4]{Tomasz Smo{\l}ka\thanks{t.smolka@uw.edu.pl}}
\affil[4]{Department of Mathematical Methods in Physics,  Faculty of Physics, University of Warsaw, Pasteura 5, 02-093, Warsaw, Poland}

\begin{document}   
\maketitle
\numberwithin{equation}{section}

\begin{abstract}
In this paper, we study some of the properties of the $G\!\to\! 0$ limit of the Kerr-Newman solution of Einstein-Maxwell equations. 
Noting  Carter's observation of the near equality between the $g=2$ gyromagnetic ratio in the Kerr-Newman solution and that of the electron, we discuss additional such coincidences relating to the Kerr-Newman multipoles and properties of the electron.
In contrast to the Coulomb field, this spinning Maxwell field has a finite Lagrangian.
Moreover, by evaluating the Lagrangian for the superposition of two such Kerr-Newman electromagnetic fields on a flat background, we are able to find their interaction potential.
This yields a correction to the Coulomb interaction due to the spin of the field. 

\end{abstract}

\section{Introduction}

\begin{figure}[htp]
    \centering
    \includegraphics[width=15cm]{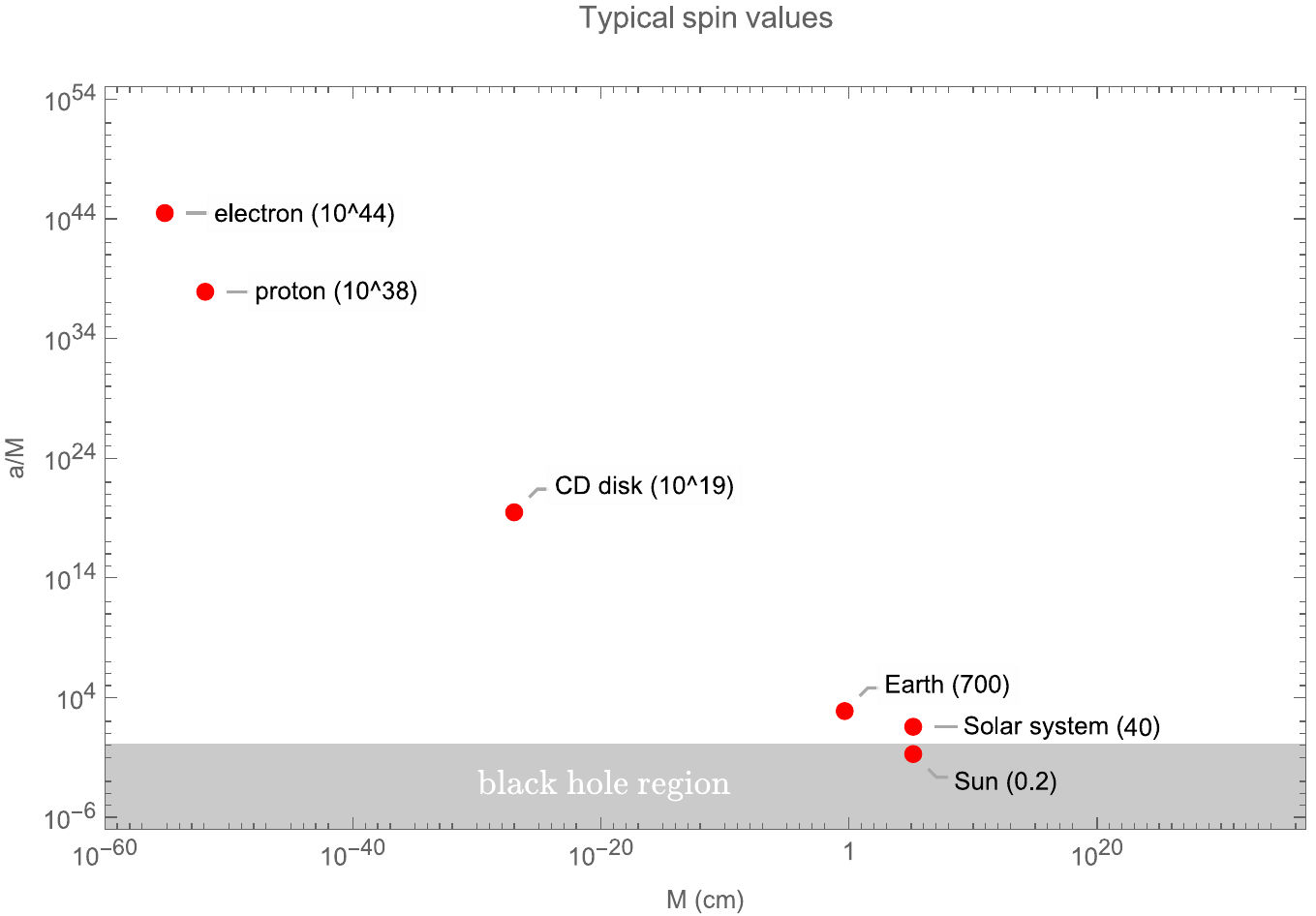}
    \caption{Values of the spin parameter $a/M$ vs. mass (by order of magnitude) for  objects with sizes ranging from the electron to the solar system. The value for a CD disk is calculated from the value $10^{18}$ for a 33 rpm vinyl LP record given by Dietz and Hoenselaers \cite{Dietz&Hoen}. The astronomical values are calculated from \cite{Allen2002}. }
    \label{fig:spin_values}
\end{figure}

This work is inspired by certain special properties of solutions
of the Einstein-Maxwell field equations of general relativity. 
As noted by Carter \cite{1968PhRv..174.1559C},
the Kerr-Newman solution has the same $g$-factor (gyromagnetic ratio) $g=2$, as the electron (apart from the
small correction to its value in quantum field theory)\footnote{The quantum field theory correction of the $g$-factor is 
$\Delta g = \alpha/2\pi \approx 0.001$ where $\alpha$ is the fine structure constant \cite{MR2148466}.}.
When comparing the electron with the Kerr-Newman solution, the quantities involved are the gravitational parameters, the mass $M$ and the spin parameter $a=J/M$ where $J$ is the angular momentum, and the electromagnetic parameters, the charge $Q$ and the magnetic moment $\mu$. For the electron, the gravitational parameters are related by the inequality $a>M$
(actually $a>>M$). This puts the comparison in the
``over-extreme'' part of the family of Kerr-Newman solutions. 
In this context, it is worth pointing out that over-extreme ($a>M$) objects are ubiquitous, ranging from elementary particles to the solar system, see Figure \ref{fig:spin_values}.

The limit $G\rightarrow0$ of the Kerr-Newman solution yields a special solution of the Maxwell equations, known as the \emph{magic field}. The term ``magic'' as coined by Lynden-Bell \cite{lyndenbell2003small} refers to the separability of motion in the field. The separability is due to the existence of a Killing tensor (of 2nd order) in the Kerr-Newman geometry 
\cite{Carter1977}. Lynden-Bell discussed some physical and mathematical properties of the magic field, in particular, whether it could be due to a charge distribution. We will instead consider the magic field as a primary entity by itself and investigate the interaction between two magic fields.

The above-mentioned coincidence of the $g$-factors
of the electron and the Kerr-Newman solution is actually a special case of related coincidences of the gravitational and electromagnetic multipoles of the Kerr-Newman solution and the known multipoles of the electron.  
The non-zero electromagnetic multipoles of the magic field coincide with those of the Kerr-Newman field. The experimentally measured values of the electron are in close agreement with five of the first ones in the two series of multipoles, electromagnetic and gravitational. They are the three first electromagnetic multipoles (charge: $q_0=Q$, electric dipole: $q_1=0$, magnetic moment: $\mu_1=Qa$) and the first and third gravitational multipoles (mass: $m_0=M$ and angular momentum: $j_1=Ma$). In particular, precise measurements of a possible electric dipole moment of the electron have so far been consistent with a zero value
\cite{andreev2018acme,2023Sci...381...46R}.

The multipole structure of the Kerr-Newman electromagnetic field consists of  alternating electric and magnetic multipoles, i.e. electric monopole (i.e.  charge), magnetic dipole, electric quadrupole etc., compare Table
\ref{E-multipoles}. The multipole structure of its gravitational field has a similar structure of alternating gravitoelectric and gravitomagnetic multipoles, i.e. gravitoelectric monopole (i.e. mass), gravitomagnetic dipole (i.e. angular momentum), gravitoelectric quadrupole etc., compare Table \ref{G-multipoles}.

It is also notable that the $g$-factor agreement depends on a certain combination of the Kerr-Newman multipoles. To see this, note that the $g$-factor is defined by the relation
\begin{equation}
    g = 2\,\frac{\Theta}{a}
\end{equation}
where $\Theta = \mu/Q$ is the magnetic dipole moment per unit charge and $a=J/m$ is the angular momentum per unit mass. Expressed in terms of multipole moments, the $g$-factor has the form
\begin{equation}
    g = 2\,\frac{m_0}{q_0} \frac{\mu_1}{j_1}
\end{equation}
where the subindices refer to the multipole order; cf. Tables \ref{E-multipoles} and \ref{G-multipoles}.

The next interesting electromagnetic multipole is the electric quadrupole, which has not been measured for the electron. 
By the Wigner-Eckart theorem  
\cite{2020mqm..book.....S}, a spin-$1/2$ particle like the electron in flat space is expected to have a vanishing electric quadrupole moment. In contrast, the quadrupole moment of the magic field is non-vanishing. As we shall see, cf. section \ref{sec:AnalyticContinuationDef} below, the magic field lives in a locally flat but topologically non-trivial spacetime, which has a ring-like, conical singularity with a $2\pi$ angle excess, giving the magic field a spin-$1/2$ aspect. In this context, it is worth pointing out that Compton found a ring-like structure for the electron \cite{1919PhRv...14..247C}. This may be taken as an indication of an oblate spheroidal structure of the electromagnetic field of the electron.
Compare also the more recent paper \cite{Koski2020}. 
The magic field originates in the Kerr-Newman solution, i.e., in a classical gravitational context, where the Wigner-Eckart theorem does not apply. Similarly, viewing the electron as a gravitating object is outside the range of validity of the Wigner-Eckart theorem, cf. \cite{Rosquist_2006}.

\begin{table}[h]
\centering
\begingroup
\setlength{\tabcolsep}{10pt} %
\renewcommand{\arraystretch}{1.5} %
\begin{NiceTabular}{cccc}[hvlines]
\RowStyle[rowcolor=lightgray]{}
l & Name & Electric $(q_{{}_l})$ & Magnetic $(\mu_{{}_l})$  \\
0 & monopole & \Frame $Q$\textcolor{red}{*} & 
\cellcolor{gray}
\\
1 & dipole & $q_{{}_1}$ & \Frame $\mu$\textcolor{red}{*}
\\
2 & quadrupole & \Frame$q_{{}_2}$ & $\mu_{{}_2}$
\\
3 & octupole & $q_{{}_3}$ & \Frame$\mu_{{}_3}$
\\
$\cdots$ & $\cdots$ & \Frame $\cdots$ & $\cdots$
\\
\end{NiceTabular} 
\endgroup
\caption{The electromagnetic multipoles $\mathcal E_{{}_l} = q_{{}_l} + i\, \mu_{{}_l}$. The non-vanishing electromagnetic multipoles of the Kerr-Newmann solution are in the cells with thick borders (yellow cells in the color version of the paper). The starred moments, i.e. the electric charge $Q$ and magnetic dipole $\mu$, are the only measured electromagnetic multipoles of the electron. The gray cell stands for a (unphysical) magnetic monopole,
that we do not consider in  this work.}

\label{E-multipoles}
\end{table}

\begin{table}[h]
\centering
\begingroup
\setlength{\tabcolsep}{10pt} %
\renewcommand{\arraystretch}{1.5} %
\begin{NiceTabular}{cccc}[hvlines]
\RowStyle[rowcolor=lightgray]{}
l & Name & \Block{}{Gravito-\\ electric $(m_{{}_l})$} & \Block{}{Gravito-\\ magnetic $(j_{{}_l})$}  \\
0 & monopole & \Frame $M$\textcolor{red}{*} & \cellcolor{gray}
\\
1 & dipole & $m_{{}_1}$ & \Frame $J \,(=Ma)$\textcolor{red}{*}
\\
2 & quadrupole & \Frame$m_{{}_2}$ & $j_{{}_2}$
\\
3 & octupole & $m_{{}_3}$ & \Frame$j_{{}_3}$
\\
$\cdots$ & $\cdots$ & \Frame $\cdots$ & $\cdots$
\\
\end{NiceTabular} 
\endgroup
\caption{The gravitational multipoles $\mathcal G_{{}_l} = m_{{}_l} + i\, j_{{}_l}$. The non-vanishing gravitational multipoles of the Kerr-Newmann solution are in cells with thick borders (yellow in the color version of the paper).  The starred moments, i.e. the mass $M$ and (spin) angular momentum $J$, are the only measured gravitational multipoles of the electron. The gray cell stands for the NUT solution, which is not relevant in the context of this work.}

\label{G-multipoles}
\end{table}

As is well known, even for the single magic field, its complete analytic continuation is not globally Minkowski. The corresponding topology has, in fact, two asymptotic ends, which are connected by two gates\footnote{Although the existence of two asymptotic ends has been discussed by a number of authors, both for the Kerr-Newman geometry and the magic field, it seems that the fact that there are two gates has not been appreciated in earlier works.}. The gates have the form of two circular disks, both of which are bounded by the (single!) ring singularity appearing in the oblate spheroidal coordinates of $R^3$. To illustrate this structure, consider a curve in one of the sheets starting from a point $P$ at one of the disks and passing around the singularity by an angle $2\pi$ to reach the other disk. To actually close the curve, one needs to pass that through that gate and pass another $2\pi$ angle on the other sheet. The curve can then be closed, having traversed an angle $4\pi$ around the singular ring. Such a curve cannot be shrunk to a point without cutting, which means that the line/curve has different topological properties than those of the curves that lie only in one of the sheets. This topological structure gives one more analogy to the electron by reference to its spin-one-half nature.

 It is worth mentioning here that the Dirac equation, which governs the relativistic dynamics of spin-$1/2$ particles, is analyzed on Kerr-Newman background by various authors \cite{Chandrasekhar1976,page1976,batic2008,Dolan2009} and is used to describe the interaction of a point electron with a structured nucleus in muonium, positronium, and hydrogen \cite{Pekeris1989}. More recently, in \cite{Gair2005} the possibility that the hyperfine splitting observed in muonium could be reproduced by the coupling between electron spin and the frame dragging caused by nuclear rotation has been examined. The Dirac equation for a point electron in a zero-G Kerr-Newman spacetime has been also studied in detail in \cite{kiessling2016dirac,kiessling2022discrete}, where the ring singularity represents the nucleus of a hydrogenic ion with charge and magnetic moment. A complete characterization of the discrete spectrum of the Dirac operator on such a background and its relation to certain topological invariants is found. In \cite{kiessling2016novel} the authors propose another perspective and reinterpret their results as for the interaction of a ring electron/positron with a point nucleus. They suggest that the radius of the ring should be associated to the anomalous magnetic moment. We don't share that perspective in this paper.

A notable fact about the magic field is that it can be obtained by making an imaginary translation of a Coulomb field in a spacelike direction. Such a procedure was described already by Appell in 1887 \cite{appell1887quelques}, cf. also \cite{Sommerfeld1896,synge1956relativity,lynden2003discs}.
This procedure is closely related to the well-known Newman-Janis algorithm \cite{newman1965note}. This involves a coordinate complexification process that transforms a static spacetime into a rotating one. Originally, the algorithm has been used to obtain the Kerr metric from the Schwarzschild and later the Kerr-Newman solution from Reissner-Nordström. 

The complex deformation process (complex shift) has been examined recently from the perspective of scattering amplitudes in \cite{arkani2020kerr}. This study analyzes the impulse imparted to a test particle by a heavy spinning particle, both in electromagnetic and gravitational contexts. In the electromagnetic case, the impulse due to the magic field (termed ``$\sqrt{\text{Kerr}}$'') is shown to be reproduced by a charged spinning particle, with the shift in the Coulomb potential corresponding to the exponentiated spin-factor in the amplitude. Based on a comparison of the result with the corresponding calculation using the scattering amplitudes, they argue that the complex shift corresponds to the exponentiation of spin operators in the large-spin limit.

Another relation between the electromagnetic magic field and the Kerr metric is through the (classical) double copy formalism, which is the origin of the nomenclature ``$\sqrt{\text{Kerr}}$'' for the magic field. In this approach, a specific class of solutions of General Relativity that admit a Kerr-Schild form introduce solutions of Maxwell theory (or non-Abelian Yang-Mills theory in a more general setting). Through this correspondence, the Schwarzschild metric is the double copy of the Coulomb solution, and the Kerr metric is the double copy of the magic field \cite{Monteiro_OConnell_White_2014}. This relation is used in \cite{arkani2020kerr} for the gravitational case to derive the impulse due to a Kerr black hole for gravitationally coupled spinning particles.

Our main object in this work has been to study interactions of magic fields at the classical level using only the magic fields themselves. This involves representing interactions in terms of localized fields rather than particles or currents. This is in contrast to the standard textbook way of dealing with classical interactions, namely, taking the particles for granted and subjecting them to fields that can be external or generated by other particles. In the textbook Lagrangian or Hamiltonian picture, the particle aspects are represented by the kinetic energy while the potentials represent the fields. Our starting point is to write down a pure field Lagrangian, which will nevertheless represent both field and particle aspects. The Lagrangian is defined as a superposition of the localized fields. These fields will, by assumption, be constructed from the magic field solutions mentioned above. Such solutions will then be valid in a flat spacetime, which is locally Minkowski but not necessarily globally Minkowski.

While our treatment is purely classical, as mentioned above, one can nevertheless make certain valid comparisons with properties of the electron and a single magic field. Turning now to the treatment of interactions of two magic fields, we have obtained effective interaction potentials for some cases. The potentials we have calculated are for aligned spins on a common axis (up-down case) and for parallel spins based on a common plane perpendicular to the spins (side-by-side case). The structure of the potentials leads to possibilities to study ``magic" type electron-electron and electron-positron interactions. In particular, in the up-down magic electron-positron case with opposite charges, there appears a state in which the fields are in equilibrium, see Figure\ref{fig:PotentialMinus}.
The state is in equilibrium with respect to the direction of the aligned spins. To know if the state is a stable equilibrium, one can compare it with the side-by-side potential in Figure \ref{fig:PotentialMinus_ss}. In that case, the potential walls indicate an attractive force when the separation of the fields is larger than the field separation in the up-down case, possibly indicating stability. Although we cannot rule out some intermediate channel between the two cases that would break stability at this stage, we consider this possibility to be unlikely. 

To compare these results with actual physical electrons and positrons, we let the magic charges equal the elementary charge. Then, in the up-down equilibrium state, the distance between equilibrium positions turns out to be equal to twice the reduced Compton wavelength, see Figure\ref{fig:Potential}.

In the following, we review the construction and essential properties of the magic electromagnetic fields in section \ref{sec:2}. In particular, we discuss the topology of the analytic extension of its potential and continuity of the field in \ref{sec:AnalyticContinuationDef} and calculate the self-Lagrangian for a single magic field in \ref{sec:self_lagr}. We analyze the interaction of two magic fields in section \ref{sec:interaction} in two specific configurations and discuss the results. Further discussions and concluding remarks are given in section \ref{sec:Conclusion}.

\section{Electromagnetic magic field}\label{sec:2}

As we mentioned in the introduction, the electromagnetic magic field can be obtained in different ways. Each of these approaches demonstrates some of the extraordinary features of this field and its relation to other solutions of physically important field equations, such as Laplace, Maxwell, and Einstein. In this section, we adopt the simplest approach to obtain the magic field from the Coulomb solution by applying a complex transformation in one of the spatial directions. We then analyze the new solution and discuss the analytic continuation of the background space on which such a solution resides. In other words, we analyze the Riemann surface and find that the magic field as a complex function is continuous and single-valued. The result is the same as the one obtained from the other approach in which one takes the limit $G\! \to\! 0$ of the Kerr-Newman solution of the Einstein-Maxwell system. In the following sections, we will review the nontrivial topology of the magic field and pave the way for the analysis of the self-energy and interaction Lagrangians.

For our analysis in this paper, it is convenient to represent the electromagnetic field by the (complex) Weber-Silberstein vector\footnote{Also known as the Riemann-Silberstein vector \cite{Bialynicki-Birula_2013}, cf. the footnote 114 in \cite{Kiessling2018}.}, 
\begin{align}\label{eq:RS_vector}
\RSB \ = \ \mathbf{E} + i \,\mathbf{B}
\end{align}
satisfying the (free) Maxwell equations
\begin{subequations}\label{eq:Maxwell's_Equations}
\begin{align}
    & \qquad \nabla \cdot \RSB \ = \ 0 \ , 
    \label{eq:Maxwell's_Equations_1}
    \\
    & i\, \partial_t \, \RSB - \nabla \times \RSB \ = \ 0 \ .
    \label{eq:Maxwell's_Equations_2}
\end{align}
\end{subequations}

Restricting to stationary\footnote{An electromagnetic field $\mathbf{F}$ is called stationary if it is preserved under the action of a time-like Killing vector field $\mathcal{T}$ of the background $(\mathcal{M}, \mathbf{g})$, i.e. $\Lie_{\mathcal{T}}\mathbf{F}=0$.} solutions, both $\mathbf E$ and $\mathbf B$ fields are curl-free  (and hence also the Weber-Silberstein vector $\RSB$) and can be obtained from scalar potentials. We introduce the complex scalar potential $\mathcal{Z}$ according to
\begin{align}\label{eq:RSB_potential}
    \RSB \ = \ - \boldsymbol\nabla \mathcal{Z} \ .
\end{align}
which generates the stationary solutions of Maxwell's equations \eqref{eq:Maxwell's_Equations}.
As a result of the (complex) Gauss equation \eqref{eq:Maxwell's_Equations_1}, the complex potential $\mathcal{Z}$ satisfies the Laplace equation
\begin{equation}
\boldsymbol\nabla^{2} \mathcal{Z} \ =  \ 0 \, .
\label{eq:field_eq_Z}
\end{equation}

The simplest potential associated with a point charge $q$ at the origin generating the Coulomb solution of Maxwell's equations is
\begin{align}\label{eq:Coulom1}
    \mathcal Z \ = \ \frac{q}R \ ,
\end{align}
where $R=\sqrt{x^2 + y^2 + z^2}$ is the distance of a given point from the origin. This potential is also proportional to the fundamental solution of the Laplacian and therefore satisfies \eqref{eq:field_eq_Z} everywhere outside the origin, where it is singular. %

Following Appell \cite{appell1887quelques} and Sommerfeld \cite{Sommerfeld1896}, we apply a complex translation $z \to z - i a$  to the Coulomb potential \eqref{eq:Coulom1} to obtain the complex potential
\begin{equation}\label{eq:magic_potential}
\mathcal Z \ = \ \frac{q}{\sqrt{\chi^{2}+(z-i\, a)^{2}}} \ ,
\end{equation}
where $\chi = \sqrt{x^2 + y^2}$ is the polar radius. Notice that we always mean the principal branch by the square root of a complex expression for which the real part is positive. 
The potential $\mathcal Z$ in \eqref{eq:magic_potential} is a complex combination of the electrostatic potential $\Phi$ and the magnetostatic potential $\Psi$ as
\begin{equation}
 \mathcal Z = \Phi + i\, \Psi \, . 
 \label{eq:magicpotential}
\end{equation}
This complex potential is, in fact, the electromagnetic (one of the pair of complex) Ernst potentials of the Kerr-Newman solution\cite{Ernst1968I,shaditahvildar2015}.
The potential $\mathcal Z$ in \eqref{eq:magic_potential} generates the electromagnetic field of the form
\begin{align}\label{eq:magic_F}
    \RSB \ = \ - \frac{q\left(\chi \, \boldsymbol{\partial}_\chi + (z- i\,a)\, \boldsymbol{\partial}_z \right)}{\left(\chi^{2}+(z-i\, a\right)^{2})^{3/2}} \ ,
\end{align}
which has both non-vanishing electric and magnetic parts, contrary to the elementary Coulomb field.

Following Lynden-Bell, \cite{lyndenbell2002magic}, we will refer to $\RSB$ and $\mathcal{Z}$ as the magic field and the magic potential, respectively. 
Similarly to its potential, the electromagnetic field is singular on the ring $\mathcal R$ at $\chi=a$ on the $z=0$ plane. This set is excluded from the domain. The field lines of electric and magnetic parts of the magic field, when viewed in Euclidean space, are displayed in Figure \ref{fig:E&B} below. The apparent discontinuity of the field on the disk spanned by the singular ring is removed by considering the analytic continuation of the field, cf. section \ref{sec:AnalyticContinuationDef}.
Under the reflection with respect to the $z=0$ plane, the electric field is symmetric, while the magnetic field is anti-symmetric.

\begin{figure}[h!]
    \centering
    \begin{subfigure}[b]{0.49\textwidth}
    \includegraphics[width=\textwidth]{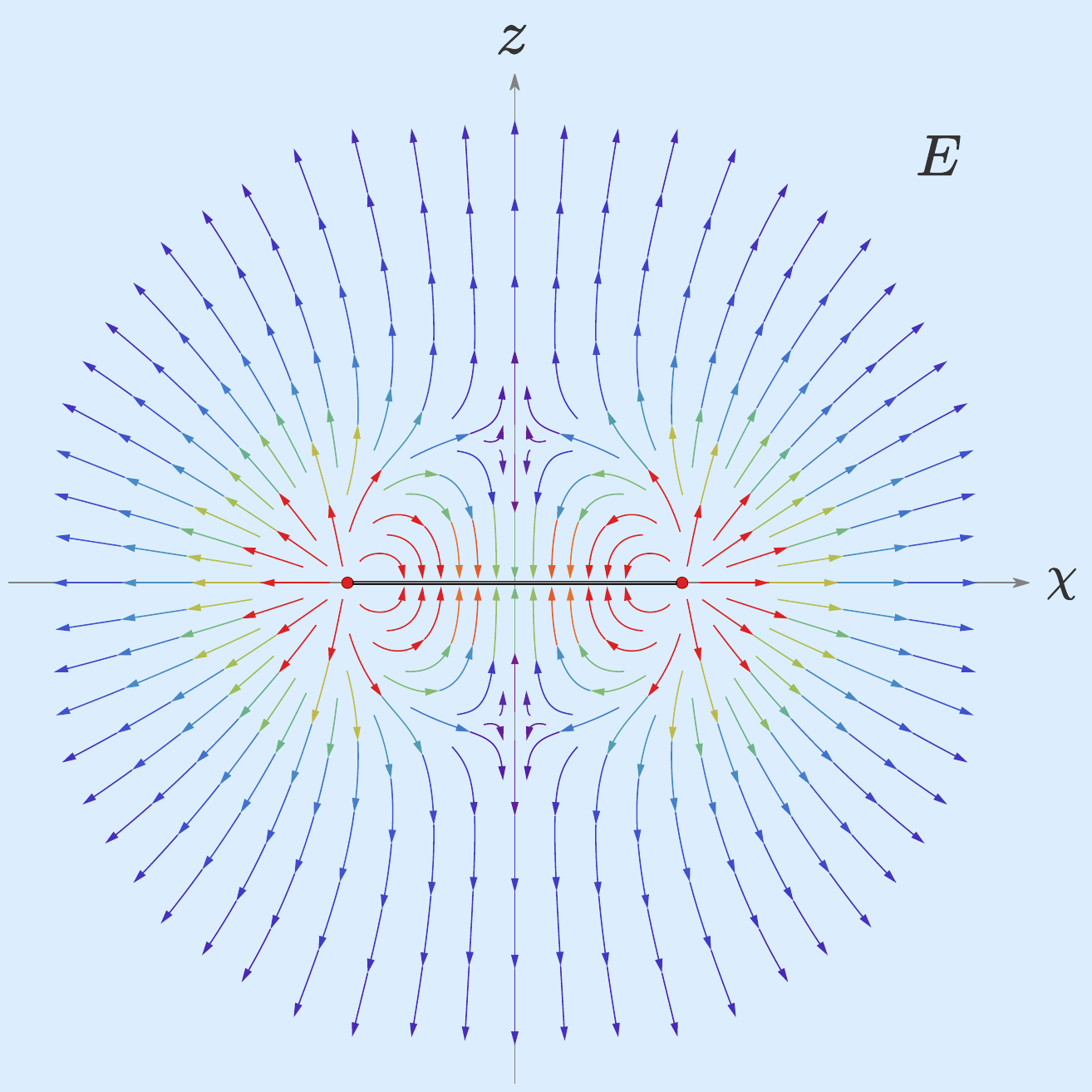}
    \caption{Electric magic field}
    \end{subfigure} \hfill%
   \begin{subfigure}[b]{0.49\textwidth}
    \includegraphics[width=\textwidth]{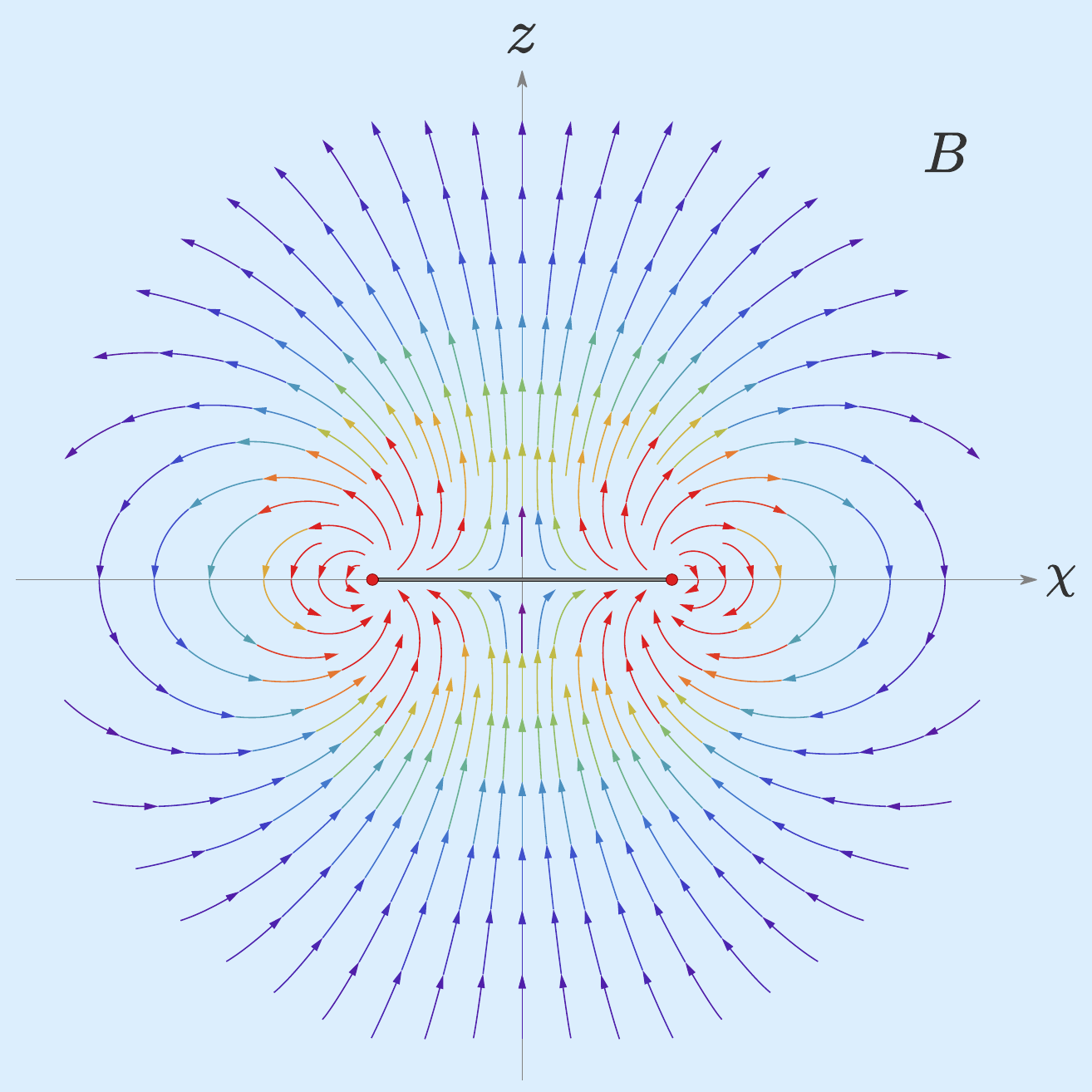}
    \caption{Magnetic magic field}
    \end{subfigure}
    \caption{Electric and magnetic components of the magic field viewed in Euclidean space. The rainbow color-coding demonstrates the fields strengths, with red representing the extreme values and purple the negligible ones. The red points on the $\chi$-axis display the singular ring $\mathcal R$ intersections, and the line segment between them displays the intersection of the branch-cut disk $\mathcal D$ spanned by the ring. The fields have azimuthal symmetry around the $z$-axis. The apparent discontinuity of the field on the disk spanned by the singular ring is removed by considering the analytic continuation of the field, cf. section \ref{sec:AnalyticContinuationDef}.}
    \label{fig:E&B}
\end{figure}

As mentioned in the introduction, the electromagnetic magic field can be obtained by taking the limit $G\! \to\! 0$ of the Kerr-Newman solution of the Einstein-Maxwell system. As a result of that limit, the background becomes locally flat with a non-trivial topology with two sheets, each having its own asymptotic region. However, the electromagnetic field of the solution remains intact in that limit, and it matches \eqref{eq:magic_F}. In the subsequent section \ref{sec:AnalyticContinuationDef}, we show how the analytic continuation 
of the potential \eqref{eq:magic_potential} extends the primary domain, i.e. the Euclidean space without the ring $\mathcal{R}$, into a two-sheeted Riemann surface. The potential is only continuous and single-valued in this extension. We also verify that the topological structure of this Riemann surface indeed coincides with that of the Cauchy surfaces of the $G\! \to\! 0$ limit of the Kerr-Newman space-time.

Another approach to explain the discontinuity of the fields at disk $\mathcal D$ taken by different researchers had been to try to find a source on the disk and the ring (see e.g. \cite{lyndenbell2003small,lynden2004electromagnetic,kaiser2004distributional}). This approach leads to exotic sources with infinite charge and current densities rotating with the speed of light. %
We do not follow this approach and will instead use  analytic continuation, thereby avoiding the problems with unphysical sources.

\subsection{Analytic continuation of the magic field}
\label{sec:AnalyticContinuationDef}
We begin our analysis by briefly discussing the complex square-root function, a lower-dimensional analog of the magic field. Recall that the complex square root function $f(\alpha)=\alpha^{1/2},$ where $\alpha$ denotes a complex variable, is double-valued over the complex plane. However, we can introduce a branch cut to make it single-valued. This process defines two separate branches of the function, which are discontinuous along the branch cut. The analytic continuation of the function across the branch cut connects two branches smoothly, resulting in a Riemann surface with two sheets glued appropriately along the branch cut. This Riemann surface, as the connected double cover of the complex plane, is the natural domain on which the square-root function is single-valued and continuous. There is no restriction on choosing the branch cut apart from the property that it connects the branch point at zero, where the value of the function agrees for both branches, to infinity. By convention, the principal branch cut is defined along the negative real axis, with the principal branch taking positive values on the positive real axis. This is our choice of branch cut in all analyses throughout this paper. For the inverse square root $\alpha^{-1/2}$ all above is valid, except that it is singular at the branch point at zero, and the Riemann surface is the connected double cover of the punctured complex plane.

For the square root in the magic field potential \eqref{eq:magic_potential}, which is a multivariable function, the set of branch points is the singular ring $\mathcal R$ at $\chi = a$ in the $z=0$ plane, and the principal branch cut is the disk\footnote{Another choice of branch cut is a surface which connects the singular ring with the infinity. However, in this case, the potential \eqref{eq:magic_potential} does not have vanishing asymptotics at infinity. In this context, the choice of branch cut on the disc $\mathcal D$ is natural.} $\mathcal D = \{\chi<a \, , \, z=0 \}$ spanned by the ring singularity, where the real part of the radicand, i.e. the argument of the square root, is negative, and its imaginary part vanishes. Therefore, the Riemann surface $\mathcal V$ of the magic potential consists of two sheets $\mathcal V_1$ and $\mathcal V_2$ glued together along the branch-cut disk $\mathcal D$. It possesses two distinct asymptotic regions, each associated with either of the sheets, and is punctured on the ring singularity. 

The values of the potential function on two sheets of its Riemann surface $\mathcal V$ are only different in overall sign. In other words, we can cover the second sheet with another copy of Cartesian coordinates so that the mapping between the two coordinate systems transforms the potential to its negative. Therefore, the potential on the whole Riemann surface in these two Cartesian coordinate systems can be written as
\begin{align}
\label{eq:magic_potential_extended}
\mathcal Z \ = \ \frac{\pm\,q}{\sqrt{\chi_n^{2}+(z_n-i\, a)^{2}}} \ , \quad n={1,2} \ ,
\end{align}
where each of the signs $\pm$ is associated with either of the sheets $\mathcal V_n$. 
In order to make the potential $\mathcal Z$ continuous along the branch-cut disk $\mathcal D$ where $\chi_n<0$ and $z_n=0$, we need to glue the upper lip of the cut from the first sheet to the lower lip of the cut from the second sheet and vice versa. The two pairs of glued lips are two distinct gates between the sheets.

The gluing of the two sheets of the Riemann surface $\mathcal V$, in the way that we explained, requires the Cartesian coordinate systems that we used in \eqref{eq:magic_potential_extended} to have opposite orientations because the flipping is along one of the coordinates (i.e., z coordinate).  We need a third coordinate system that covers the neighborhood around the branch cut in both sheets and overlaps with the two Cartesian coordinate systems to explore this and other nontrivial topological features. %
In the following, we will use the oblate spheroidal coordinates, extended by analytic continuation for this purpose.

The Riemann surface $\mathcal V$ of the magic potential that was described earlier shares the topology of spatial hypersurfaces of the spacetimes discussed by Zipoy \cite{Zipoy1966} and matches that of the $m\!\to\!0$ limit of the maximal analytic continuation of the Kerr solution and $G\! \to\! 0$ limit of the Kerr-Newman solution \cite{1968PhRv..174.1559C,shaditahvildar2015,Gibbons-Volkov2016,Gibbons-Volkov2017}. This will be most evident when we introduce the oblate spheroidal coordinates to our analysis. The oblate spheroidal coordinates $(r,\theta,\phi)$ are defined via the transformation
\begin{align}
x \ = \ \sqrt{r^2+a^2}\, \sin \theta\,\cos\phi \ , \quad y \ =  \ \sqrt{r^2+a^2}\, \sin \theta\,\sin\phi \ ,
\quad z \ & = \ r \cos \theta \ ,
\label{eq:OblateSpheroidalCoordCart}
\end{align}
in which we have $\chi=\sqrt{r^2+a^2}$. Note that the oblate coordinates $r$ and $\theta$ are different from the standard spherical radial and angular coordinates. Their level sets are, respectively, the spheroids and hyperboloids of revolution around the $z$-axis. Here $r$ is the length of the semi-major axis of spheroids with their foci at the ring $\chi=a$ on the $z=0$ plane and $\theta$ is the angle of asymptotes of the hyperboloid. On the principal branch, where we first obtained the magic potential \eqref{eq:magic_potential}, the oblate $r$-coordinate is a non-negative real number and $0\leq\theta \leq\pi\,,\, 0\le\phi\le2 \pi $. We will see that the extension of the $r$-coordinate to negative values enables the oblate coordinate system to cover the whole Riemann surface in a beneficial way.

The oblate spheroidal coordinates adapt very well to the singularity and topological structure of the Riemann surface of the magic field. The ring singularity $\mathcal R$ of the magic potential \eqref{eq:magic_potential}, which is the branch point set of the square root, coincides with the focal ring of $r$-constant spheroids and the coordinate singularity at $\{r=0,\,\theta=\pi/2\}$. The principal branch cut at the disk $\mathcal D$ spanned by the ring $
\mathcal R$ is mapped into two separate disks at $\mathcal D_1 = \{r=0,\,0\leq\theta<\pi/2\}$ and $\mathcal D_2=\{r=0,\,\pi/2<\theta\leq\pi\}$. These are the two separate lips of the branch cut $\mathcal D$ with the same boundary at the ring $\mathcal R$, which are to be glued ``correctly" to their counterparts from the second sheet after the extension that we perform. We then denote the glued lips with $
\mathcal D_1$ and $
\mathcal D_2$ again, for obvious reasons, and call them the ``gates" between the two sheets of the Riemann surface $\mathcal V$.

The analytic continuation that leads to the whole Riemann surface of the potential is the most straightforward in oblate spheroidal coordinates and needs only the continuation of the oblate $r$ coordinated to include negative values to cover the second sheet. Therefore, the equations \eqref{eq:OblateSpheroidalCoordCart} give at once the transformation from both copies of the Cartesian (polar) coordinates used in \eqref{eq:magic_potential_extended}. The big advantage of this is that the gluing of the sheets together is done correctly, as we need for the continuity of the potential over the branch cut lips (gates).

The validity of the coordinate transformation \eqref{eq:OblateSpheroidalCoordCart} on the whole Riemann surface $
\mathcal V$, enables us to write the analytically continued magic potential $\mathcal Z$ in \eqref{eq:magic_potential_extended} in oblate spheroidal coordinates as 
\begin{equation}
\mathcal Z \ = \ \frac{\pm\,q}{\sqrt{(r- i a \cos \theta)^{2}}} \ = \ 
\frac{q}{r- i a \cos \theta}
\, ,
\label{eq:PotentialOblate}
\end{equation}  
where in the last equality, the different signs from the square root simplification in either of sheets with $r \ge 0$ or $r \le 0$ cancels the sign associated with the relevant sheet, and the last expression holds for the entire Riemann surface. The merit of exploiting the oblate spheroidal coordinates in covering the whole Riemann surface of the magic potential becomes evident again in the fact that the last simple result in \eqref{eq:PotentialOblate} is obviously continuous on the two gates $\mathcal D_1 = \{r=0,\,0\leq\theta<\pi/2\}$ and $\mathcal D_2=\{r=0,\,\pi/2<\theta\leq\pi\}$. The coordinate regions of the maximal analytic continuation of the Riemann surface $\mathcal V$ of the magic potential in extended oblate spheroidal coordinates are listed in Table \ref{Magic_coords}.
\begin{table}[htp]
\centering
\begingroup
\setlength{\tabcolsep}{10pt} %
\renewcommand{\arraystretch}{1.5} %
\begin{tabular}{ |c|c|p{6cm}|  }
\hline
\rowcolor{lightgray} \multicolumn{3}{|c|}{ Topology of the Riemann Surface $\mathcal V$}  \\
\hline
Sheet 1 & $\mathcal V_1$ %
& 
\begin{tabular}{@{}c@{}}
$\mathcal V_1^+ \ = \ \{r\geq0,\  0\leq\theta\leq\pi/2\}$
\\
$\mathcal V_1^- \ =\ \{r\geq0,\  \pi/2<\theta\leq\pi\}$
\end{tabular}
\\
\hline
Sheet 2 & $\mathcal V_2$ %
& 
\begin{tabular}{@{}c@{}}
$\mathcal V_2^+ \ = \ \{r\leq0,\ \pi/2\leq\theta\leq\pi \}$
\\
$\mathcal V_2^- \ =\ \{r\leq0,\ 0\leq\theta<\pi/2 \}$
\end{tabular}
\\
\hline
Gate 1 & $\mathcal D_1$ &
$ \mathcal V_1^+\cap \mathcal V_2^- \ = \ \{r=0,\  0\leq\theta<\pi/2\}$
\\
\hline
Gate 2 & $\mathcal D_2$ &
$\mathcal V_1^-\cap \mathcal V_2^+ \ = \ \{r=0,\  \pi/2<\theta\leq\pi\}$
\\
\hline
Ring Singularity & $\mathcal R$ &
$\partial\mathcal D_1 \ = \ \partial\mathcal D_2 \ = \ \{r=0,\  \theta=\pi/2\}$
\\
\hline
\end{tabular}
\endgroup
\caption{Topology of the maximal analytic continuation of the magic potential and the coordinate regions of its constituents. The signs $\pm$ stand for the ``upper-half" ($z\geq0$) and ``lower-half" ($z\leq0$) parts of the sheets. By convention, we have included the gate parts, $r=0$, in sheet definitions for easy specification of the two gates as intersections of the two sheets.}
\label{Magic_coords}
\end{table}

\begin{figure}[htp]
    \centering
    \begin{subfigure}{0.49\textwidth}
        \centering
        \includegraphics[width=\textwidth]{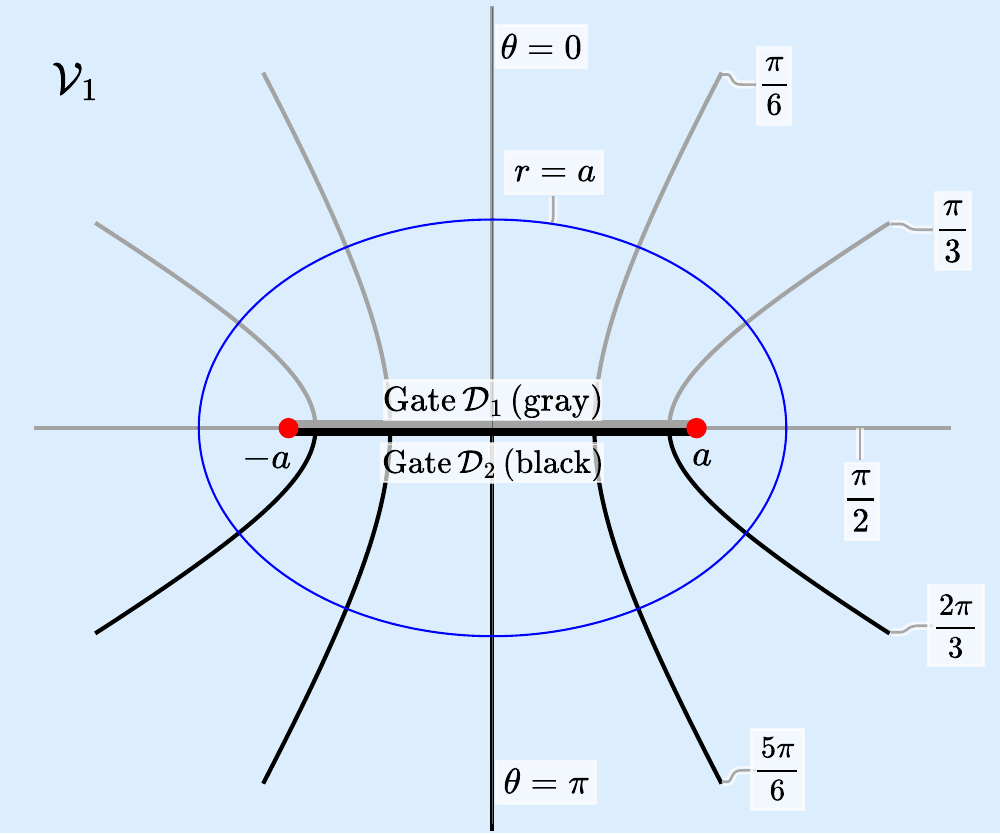} %
        \caption{First sheet $\mathcal V1$}
        \label{fig:FirstSheet}
    \end{subfigure}\hfill
    \begin{subfigure}{0.49\textwidth}
        \centering
        \includegraphics[width=\textwidth]{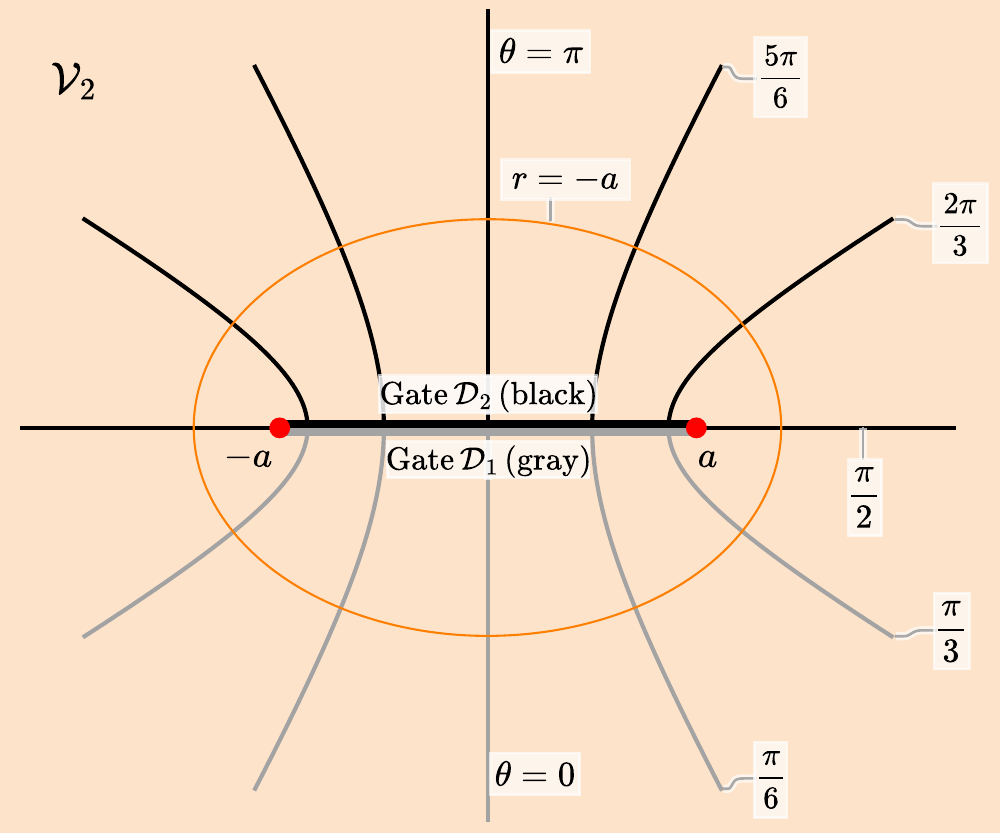} %
        \caption{Second sheet $\mathcal V_2$}
        \label{fig:SecondSheet}
    \end{subfigure}
    \par\bigskip
\begin{subfigure}{0.49\textwidth}
        \centering
        \includegraphics[width=\textwidth]{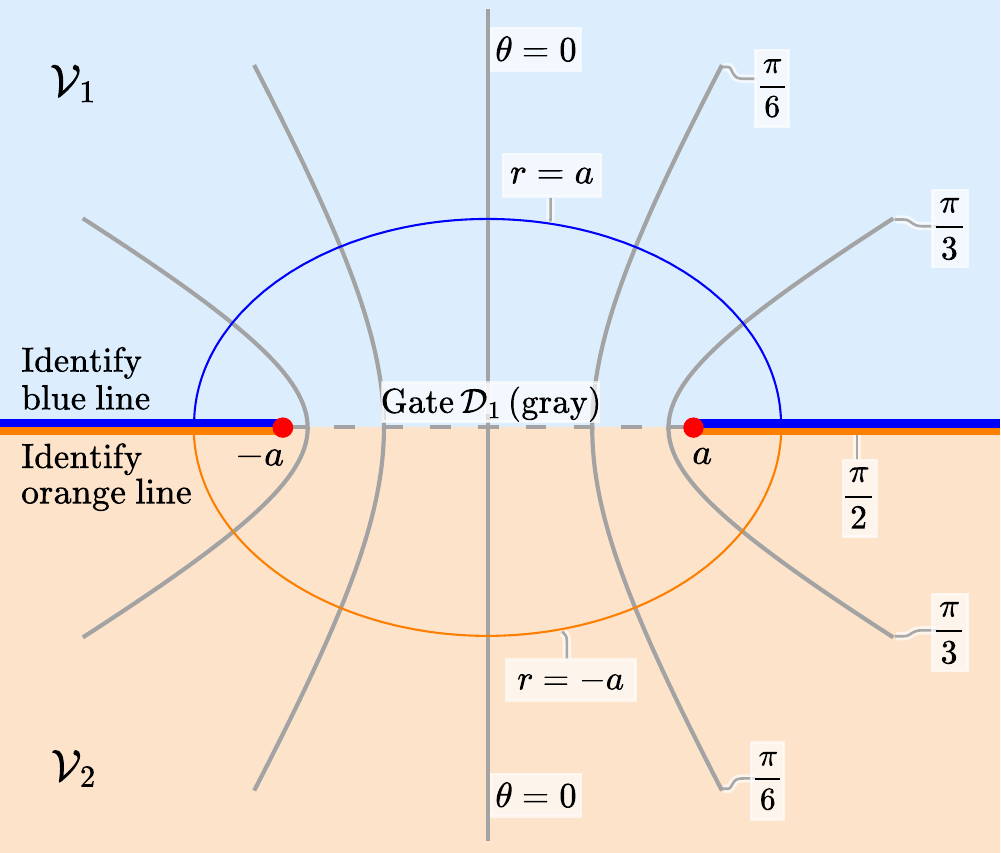} %
        \caption{First gate $\mathcal D1$}
        \label{fig:FirstGate}
    \end{subfigure}\hfill
    \begin{subfigure}{0.49\textwidth}
        \centering
        \includegraphics[width=\textwidth]{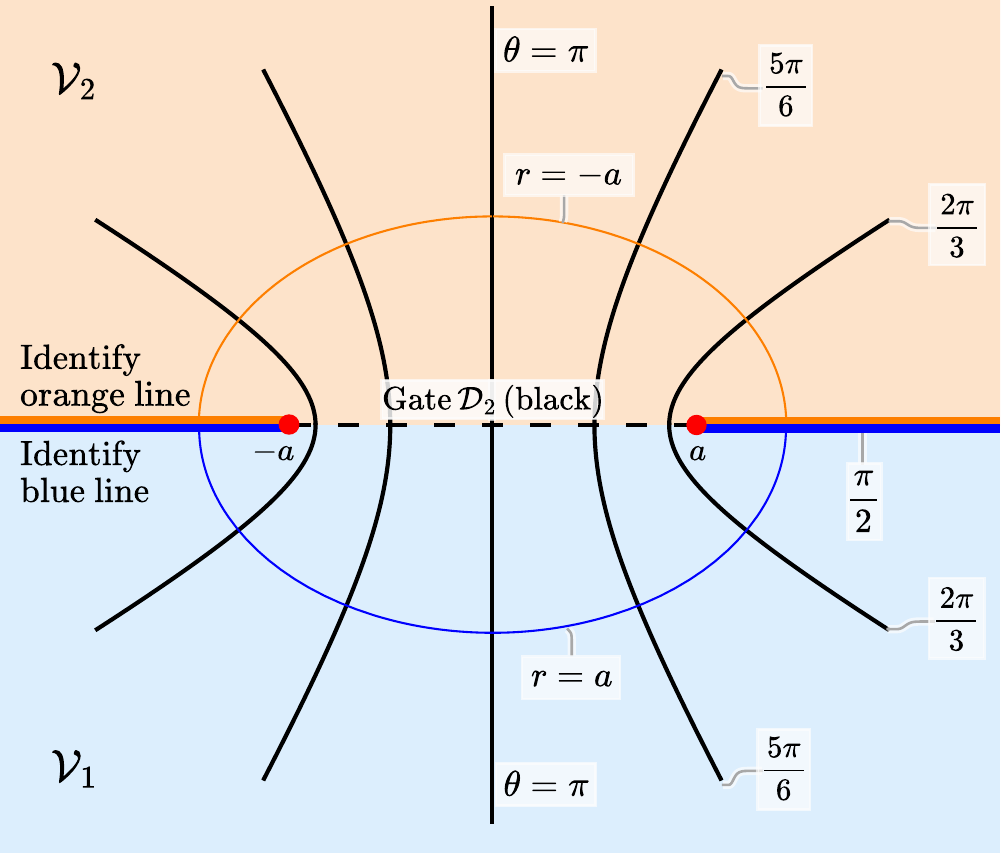} %
        \caption{Second gate $\mathcal D_2$}
        \label{fig:SecondGate}
    \end{subfigure}
    \caption{The whole Riemann surface $\mathcal V$ of the electromagnetic magic field: (a) the first sheet $\mathcal V_1$, (b) the second sheet $\mathcal V_2$, (c) the first gate $\mathcal D_1$ between the upper part of the first sheet $\mathcal V_1^+$ and the lower part of the second sheet $\mathcal V_2^-$, and (d) the second gate $\mathcal D_2$ between the lower part of first sheet $\mathcal V_1^-$ and the upper part of the second sheet $\mathcal V_2^+$. The red points represent the ring singularity $\mathcal R$. The blue and orange line segments that separate the upper and lower halves in (c) and (d) are identified respectively. }
    \label{fig:Sheets&Gates}
\end{figure}

\begin{figure}[htp!]
    \centering
    \begin{subfigure}[b]{0.49\textwidth}
    \includegraphics[width=\textwidth]{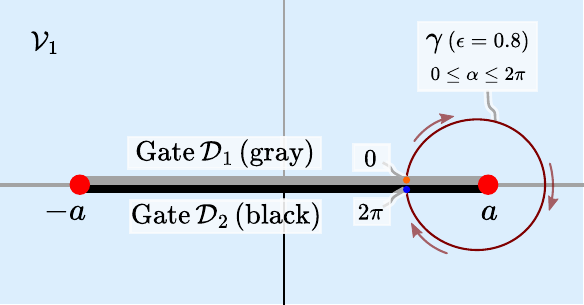}
    \caption{First round $0\le \alpha \le 2\pi$.}
    \end{subfigure} \hfill
   \begin{subfigure}[b]{0.49\textwidth}
    \includegraphics[width=\textwidth]{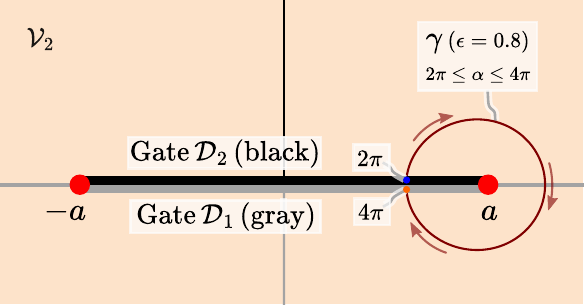}
    \caption{Second round $2\pi\le \alpha \le 4\pi$.}
    \end{subfigure}
    \caption{Curve $\gamma$ circling around the ring singularity $\mathcal R$ in double-sheeted Riemann surface $\mathcal V$. For  $0\le\alpha\le2\pi$, the curve $\gamma$ is in the first sheet $\mathcal V_1$ starting from the orange point $\alpha=0$. It enters the second sheet $\mathcal V_2$ from the black gate $\mathcal D_2$ at $\alpha = 2\pi$ depicted by the blue point on the gate. For $2\pi\le\alpha\le4\pi$, it is in the second sheet $\mathcal V_2$ and enters back to the first sheet from the gray gate $\mathcal D_1$ and closes at the starting point. The depicted curve is an $\epsilon$-constant curve, $\epsilon$ being the semi-radius of the curve defined in \eqref{eq:ring_centered_coords1}.}
    \label{fig:4pi_curve}
\end{figure}

Under the transformation \eqref{eq:OblateSpheroidalCoordCart}, the background metric transforms to
\begin{align}\label{eq:metric}
    \ud s^2 \ = \ \frac{\Sigma}{r^2+a^2}\,\ud r^2 + \Sigma\, \ud\theta^2 + (r^2+a^2)\sin^2\theta\,\ud\phi^2
\end{align}
where $\Sigma = r^2 + a^2 \cos^2\theta$. This is the $m\!\to\!0$ limit of the Kerr metric and the $G\! \to\! 0$ limit of the Kerr-Newman metric. The metric \eqref{eq:metric} is flat everywhere outside the ring $\mathcal R$. A consequence of the nontrivial topology of the Riemann surface $\mathcal V$ is that if we round the ring $\mathcal R$ and pass through the disk $\mathcal D$, say in a $\phi$-constant plane, we need to go round twice to return to the starting point. For more explanation, see the following discussion and also Figure \ref{fig:4pi_curve}. This $4\pi$ angle rounding means that $\mathcal R$ is a conical singularity with a $2\pi$ angular excess. 

The $2\pi$ angular excess can be demonstrated by introducing the closed curve  $\gamma$ rounding the ring $\mathcal R$ on $\phi$-constant plane parameterized by an angle $\alpha$ so that
\begin{align}\label{eq:ring_centered_coords1}
    r \ = \ \epsilon\, |a|\, \sin \frac\alpha2 \quad, \quad\cos\theta \ = \ \epsilon\,\cos\frac\alpha2 \ .
\end{align}
For  $0<\alpha<2\pi$, the curve $\gamma$ circles the ring in the first sheet $\mathcal V_1$, and for $2\pi<\alpha<4\pi$, it circles that in the second sheet $\mathcal V_2$, cf. Figure \ref{fig:4pi_curve}. 

For a small $\epsilon$, the shape of $\gamma$ tends to become a circle with the circumference as 
\begin{align}
    \int_0^{4\pi} \sqrt{g(\dot\gamma,\dot\gamma)} \,\ud\alpha \ = \ 2\pi\, |a|\, \epsilon^2 + \mathcal O(\epsilon^3)
\end{align}
and the radius, i.e. the distance from its center on the ring $\mathcal R$, as
\begin{align}
    \sqrt{(\chi - a)^2 + z^2} \ = \ \frac{|a|}2\,\epsilon^2 + \mathcal O(\epsilon^3) \ .
\end{align}
Therefore, the ratio of the circumference to the radius is $4\pi$ in the limit that $\gamma$ shrinks to a point. The same analysis is done in \cite{shaditahvildar2015} by a slightly different parametrization. We will use the relations \eqref{eq:ring_centered_coords1} in introducing a useful coordinate system later.

\begin{figure}[htp]
    \centering
    \begin{subfigure}[b]{0.49\textwidth}
    \includegraphics[width=\textwidth]{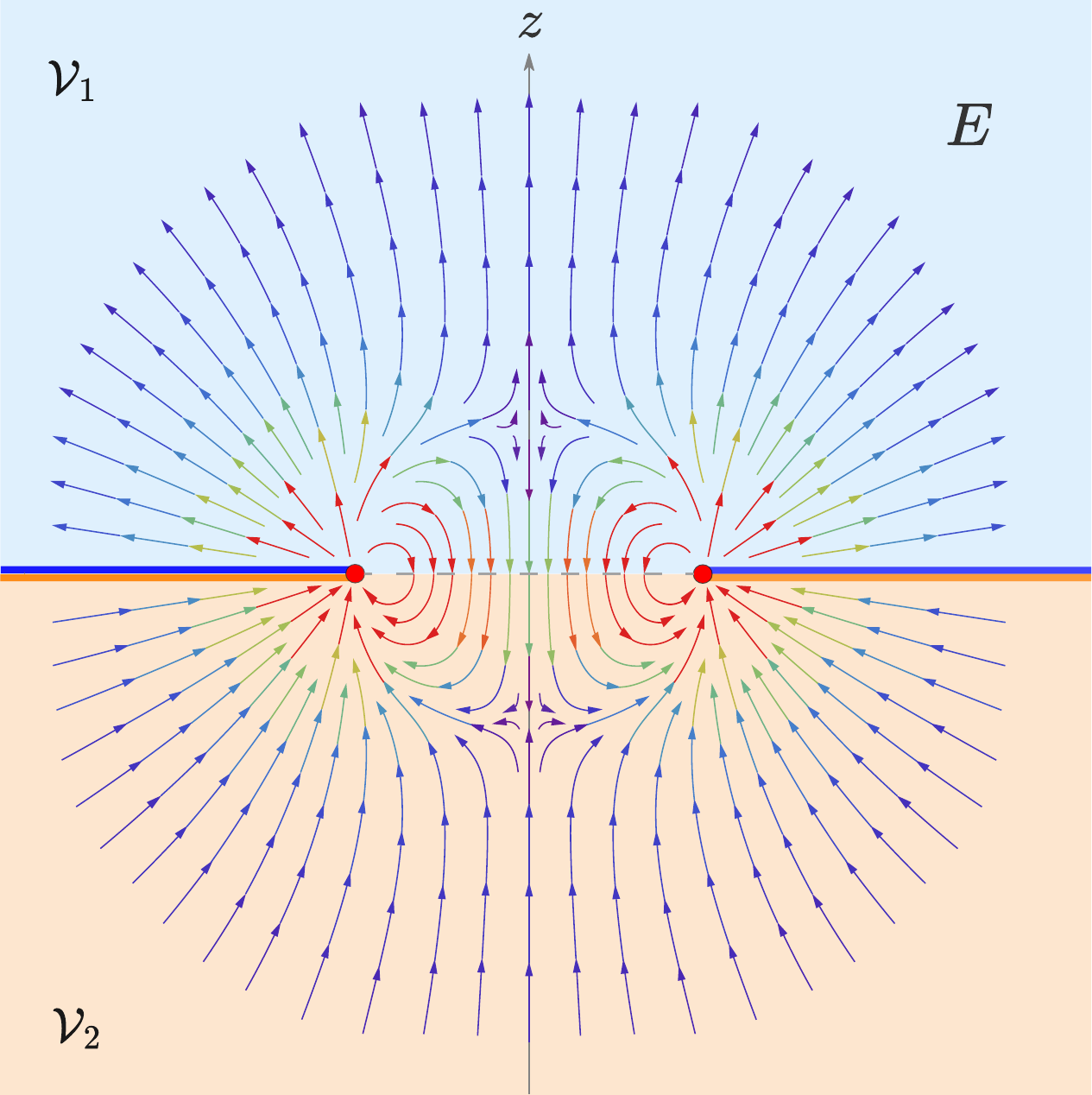}
    \caption{Electric magic field}
    \end{subfigure} \hfill
   \begin{subfigure}[b]{0.49\textwidth}
    \includegraphics[width=\textwidth]{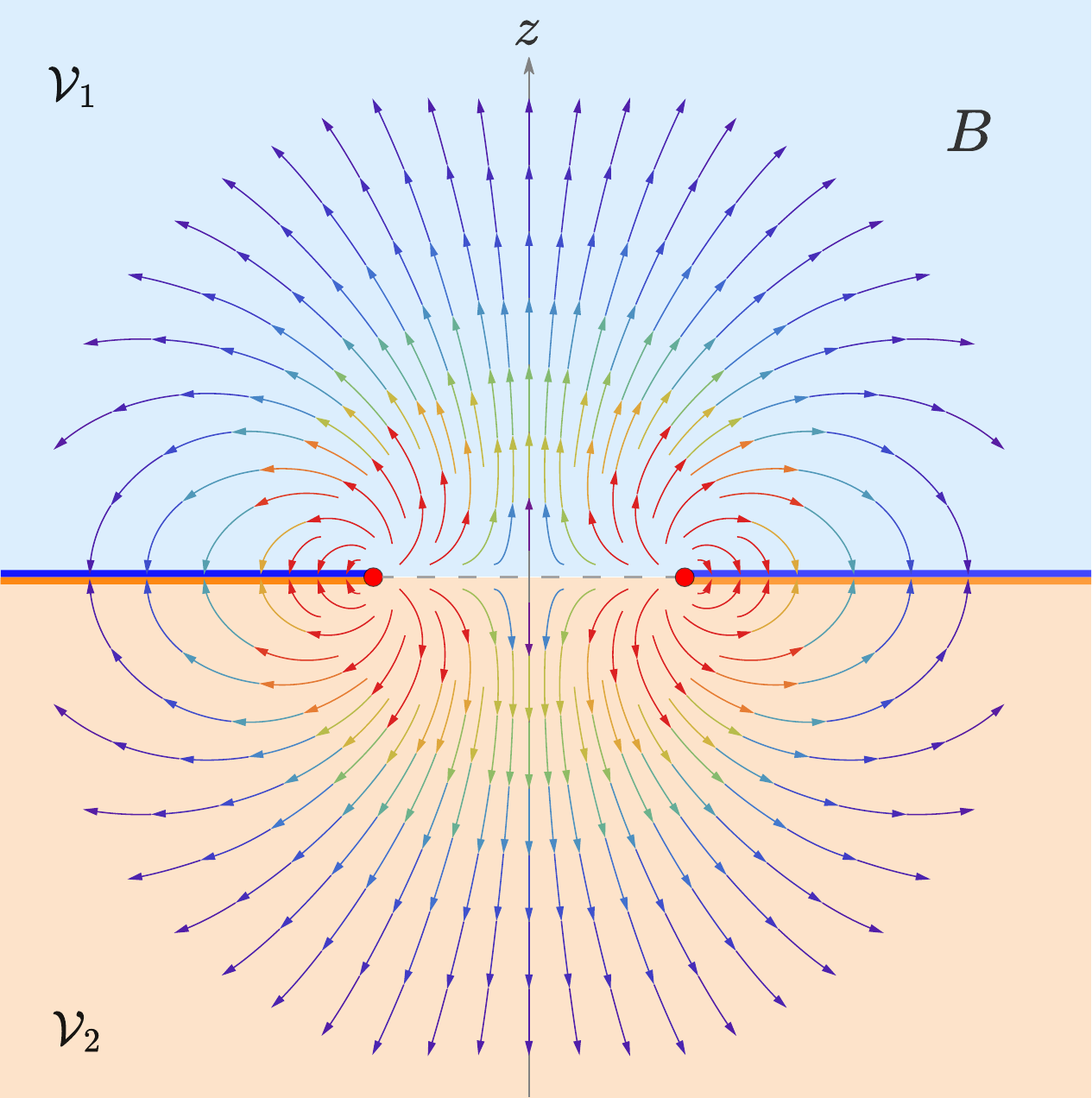}
    \caption{Magnetic magic field}
    \end{subfigure}
    \caption{Electric and magnetic components of the magic field through the gate $\mathcal D_1$ of the Riemann surface $\mathcal V$ of the complex potential. The rainbow color-coding demonstrates the fields strengths, with red representing the extreme values and purple the negligible ones. 
    The red points on the $\chi$-axis display the singular ring $\mathcal R$ intersections, and
    the line segment between them displays the intersection of the branch-cut disk $\mathcal D$ spanned by the ring $\mathcal R$. The fields have azimuthal symmetry around the $z$-axis. There is no discontinuity of the field on the disk spanned by the singular ring. Similar graphs can show the fields structure in the other gate $\mathcal D_2$.}
    \label{fig:E&B_2sheet}
\end{figure}

It is noteworthy that by considering the analytic continuation, the apparent discontinuity of the electric and magnetic fields in Figure \ref{fig:E&B} is no longer present, cf. \ref{fig:E&B_2sheet}. 

\subsection{Self-energy and Lagrangian of a magic field}\label{sec:self_lagr}

In this section, we analyze the free Electromagnetic Lagrangian for a single magic field in its nontrivial two-sheeted background topology. To this end, we exploit the stationarity of the magic field to write the electromagnetic Lagrangian integral over the spatial space as a surface integral using Stokes' theorem. The method developed here will be applied to our analysis of interaction Lagrangian of the magic fields in the following sections.

\paragraph{Electromagnetic Lagrangian}
Let $\mathcal V$ be a spatial leaf (i.e. a Cauchy time slice) of the spacetime manifold $\mathcal M$. The total (free) electromagnetic Lagrangian, in terms of the Weber-Silberstein vector \eqref{eq:RS_vector} can be written as
\begin{align}
L \ & = \ \frac{1}{2\, \Omega} \int_{\mathcal V} \left(\mathbf{E}^2 - \mathbf{B}^2\right) \ud V \ = \ \Re{[\mathcal L]} \ ,
\label{MaxLagr}
\end{align}
where $\mathcal L$ is the complex Lagrangian\footnote{We always use this complex Lagrangian throughout the text, and the standard Lagrangian is just the real part of it, see Eq. \eqref{MaxLagr}.} defined as
\begin{align}
\mathcal L \ & := \ \frac{1}{2\, \Omega}  \int_{\mathcal V} \RSB^{2}\, \ud V \ . 
\label{MaxLagr1}
\end{align}
In the above expressions  $\ud V$ is the volume element on the spatial leaf $\mathcal V$, and $\Omega$ is the solid angle of the whole spatial infinity of the asymptotic region of $\mathcal V$. In usual cases, where ${\mathcal V}$ is one-sheeted, $\Omega = 4\pi$. However, in the case of the magic field, where the background is two-sheeted, we have $\Omega = 8\pi$. Although it is unnecessary to specify the value of $\Omega$ for the analysis in this section, we will see later that it becomes important to recover the Coulomb limit of the interaction consistently.

Using the potential representation for stationary solutions \eqref{eq:RSB_potential}, the complex Lagrangian \eqref{MaxLagr1} is recast in
\begin{eqnarray}
\mathcal L &={}&\frac{1}{2\, \Omega}\, \int_{\mathcal V} \RSB^{2}\, \ud V
\ = \ \frac{1}{2\, \Omega}\, \int_{\mathcal V} \boldsymbol\nabla \mathcal{Z} \cdot \boldsymbol\nabla \mathcal{Z} \,  \ud V
\nonumber
\\
&={}&\frac{1}{2\, \Omega}\, \int_{\mathcal V}
\left[\boldsymbol\nabla \cdot(\mathcal{Z} \boldsymbol\nabla \mathcal{Z})-\mathcal{Z}\, \boldsymbol\nabla^{2} \mathcal{Z}\right] \ud V
\nonumber
\\
&={}&
\frac{1}{4\, \Omega} \oint_{ \partial \mathcal V} \boldsymbol\nabla \mathcal{Z}^2 \cdot \mathrm{d} \boldsymbol S
\, ,
\label{eq:Lagrangian_in_Z}
\end{eqnarray}
where $\mathrm{d} \boldsymbol S$ is the directed area element of the boundary $\partial \mathcal V$ toward outside, and for the last equality, we have used Stokes' theorem and used \eqref{eq:field_eq_Z} on $\mathcal V$.

\paragraph{Lagrangian of a single magic field} To analyze the Lagrangian \eqref{eq:Lagrangian_in_Z} for a single magic field on its Riemann surface $\mathcal V$, we need to locate the boundary $\partial\mathcal V$. The outer boundary consists of two distinct asymptotic regions, where the value of the potential and field tends rapidly enough to zero. The inner boundary is a double torus that surrounds the ring singularity from both sheets. We can describe this surface by the limit of a level surface of $\Sigma$, when its value tends to zero. If we use the equations \eqref{eq:ring_centered_coords1} to introduce a new coordinate system $(\epsilon,\alpha,\phi)$ as
\begin{align}
    r= \epsilon\, |a|\, \sin \frac\alpha2 \quad, \quad\cos\theta = \epsilon\,\cos\frac\alpha2 \quad, \quad \phi = \phi \ .
\end{align}
In this coordinate system, we have
\begin{align}
    \Sigma \ = \ r^2 + a^2 \cos^2\theta \ = \ a^2\,\epsilon^2 \ .
\end{align}
The non-orthonormal basis vectors in this coordinate system consist of
\begin{align}
    \boldsymbol\partial_\epsilon \ & = \ \frac r\epsilon \,\boldsymbol\partial_r - \frac{\cot\theta}\epsilon \,\boldsymbol\partial_\theta
\\
    \boldsymbol\partial_\alpha \ & = \ \frac {|a|}{2}\,\cos\theta \, \boldsymbol\partial_r + \frac r{2 |a|}\,\csc\theta\,\boldsymbol\partial_\theta
\end{align}
and $\boldsymbol\partial_\phi$. The (outward) unit normal vector to the surface reads
\begin{align}
    \boldsymbol{n} \ = \ \frac{\sqrt{1-\epsilon^2\,\cos\alpha}}{|a|\,\epsilon}\,\boldsymbol\partial_\epsilon + \frac{\sin\alpha}{|a|\,\sqrt{1-\epsilon^2\,\cos\alpha}}\,\boldsymbol\partial_\alpha \ .
\end{align}
Using $(\alpha,\phi)$ parameterization of a $\epsilon$-constant surface, the (outward) vectorial area element can be written in terms of the tangent vectors $\boldsymbol\partial_\alpha$ and $\boldsymbol\partial_\phi$ as
\begin{align}\label{eq:surface_element_2}
    (\ud \boldsymbol S)^i \ = \ \sqrt{g}\, \varepsilon^i{}_{jk}\,(\boldsymbol\partial_\alpha)^j \, (\boldsymbol\partial_\phi)^k \ud \alpha\ud\phi \ ,
\end{align}
where $\sqrt g = |a|^3\,\epsilon^3/2$ and $\varepsilon_{ijk}$ is the Levi-Civita symbol in new coordinates\footnote{We can equivalently find the area element as $\ud \boldsymbol S = \sqrt{h}\,\boldsymbol n$, with $h$ being the determinant of the induced metric on the toroidal $\Sigma$-surface, which leads to the same result.}

The Lagrangian of the magic field on its Riemann surface $\mathcal V$ is obtained as
\begin{align}\label{eq:single_magic_Larangian}
    \mathcal L \ & = \ \frac{1}{4\Omega}\oint_{\partial \mathcal V} \boldsymbol\nabla \mathcal Z^2 \cdot \ud \boldsymbol S 
\nonumber \\
& = \ \frac1{32\pi}\int_0^{2\pi} \!\!\!\ud\phi \int_0^{4\pi} \!\!\!\ud \alpha \, \frac{q^2}{4\,|a|}\left(1+3\,e^{-2i\alpha} - 4\,\epsilon^{-2}\,e^{-i\alpha}\right)
\nonumber\\
& = \ \frac{\pi \,q^2}{16\, |a|} \ .
\end{align}
Note that here, the integral over the outer boundary at infinity vanishes, and the volume form on the inner toroidal surface is inward, thus a minus sign with respect to \eqref{eq:surface_element_2}. The value of Lagrangian for the magic field given by \eqref{eq:single_magic_Larangian} is therefore finite for a finite spin parameter $a$ and diverges in the Coulomb limit $a\!\to\!0$ as expected. In the same manner, we can calculate the self-energy, which is obtained as
\begin{align}
    \mathcal E \ & = \ \frac{1}{4\Omega}\oint_{\partial \mathcal V} \boldsymbol\nabla (\mathcal Z \mathcal Z^*) \cdot \ud \boldsymbol S 
\nonumber \\
& = \ \lim_{\epsilon\to0}\,\frac{q^2}{16\, |a|\,\epsilon^2} \ ,
\end{align}
which diverges, as is the case for the Coulomb field.

\section{Interaction of two magic fields} \label{sec:interaction}
In this section, we study the interaction of two electromagnetic magic fields. 
We 
calculate the interaction Lagrangian for two Coulomb fields, as well as for two electromagnetic magic fields in two cases: first, in an up-down configuration when their symmetry axis coincides, and second, in a side-by-side configuration when their ring singularities are laid in the same plane. We will discuss the topology of the superposition of magic fields in these configurations.

Since Maxwell's equations \eqref{eq:Maxwell's_Equations} are linear equations, the superposition of two electromagnetic fields
$
\RSB_{\text{tot}} = \RSB_1 + \RSB_2 
$
is a solution of those equations if its ingredients 
are so. 
For such a superposition, the total (complex) Lagrangian \eqref{MaxLagr1} is recast into
\begin{align}\label{eq:tot_Lagrangian}
\mathcal L_{\text{tot}} \ = \ \mathcal L_1 + \mathcal L_2 + \mathcal L_{\text{int}}  \ ,
\end{align}
with definitions of (complex) \textit{self-Lagrangians} as
\begin{align}
    \mathcal L_n \ = \ \frac{1}{2\,\Omega} \int_{\mathcal V} \RSB_n^2 \ \ud V \quad , \quad n=1,2\ ,
\end{align}
for either of the fields\footnote{In the Coulomb case, the self-Lagrangian diverges, while for the magic field, the integral is convergent; see \eqref{eq:single_magic_Larangian}.}, and the (complex) \textit{interaction Lagrangian} as
\begin{align} \label{eq:int_Lagrangian}
\mathcal L_{\text{int}} \ = \ \frac{1}{\Omega} \int_{\mathcal V} \RSB_1 \cdot \RSB_2 \ \ud V \, .
\end{align}
In the static case, where the kinetic term is absent, the interaction Lagrangian can be viewed as defining an interaction potential. 

For stationary solutions $\RSB_1 = - \boldsymbol\nabla \mathcal Z_1$ and $\RSB_2 = - \boldsymbol\nabla \mathcal Z_2$, the interaction Lagrangian \eqref{eq:int_Lagrangian} is
\begin{align} \label{eq:int_Lagrangian_potentials}
\mathcal L_{\text{int}} \ & = \ \frac{1}{\Omega} \int_{\mathcal V} \boldsymbol\nabla \mathcal Z_1 \cdot \boldsymbol\nabla \mathcal Z_2 \, \ud V \nonumber \\
\ & = \ \frac{1}{2\,\Omega}  \int_{\mathcal V} \left[ \boldsymbol\nabla \cdot ( \mathcal Z_1 \, \boldsymbol\nabla \mathcal Z_2 + \mathcal Z_2 \, \boldsymbol\nabla \mathcal Z_1) - \mathcal Z_1 \, \boldsymbol\nabla^2 \mathcal Z_2  - \mathcal Z_2 \, \boldsymbol\nabla^2 \mathcal Z_1 \right] \ud V \nonumber \\
& = \ \frac{1}{2\,\Omega}  \oint_{ \partial \mathcal V} \boldsymbol \nabla (\mathcal{Z}_1 \mathcal{Z}_2) \cdot \mathrm{d} \boldsymbol S
\end{align}
where $\mathcal V$ is the Riemann surface of $\mathcal F_{\text{tot}}$ and the last equality holds when the divergence 
terms vanish inside $\mathcal V$ 
and we take the limit toward the singular regions for the boundary $\partial \mathcal V$.

For the superposition of two Coulomb fields, the self-Lagrangians $\mathcal L_1$ and $\mathcal L_2$ in \eqref{eq:tot_Lagrangian} diverge. However, the interaction Lagrangian \eqref{eq:int_Lagrangian} is still well-defined and can be evaluated as follows. Consider two point charges located at $z=\pm z_0$
\begin{align}\label{eq:Coulomb_field}
\mathcal Z_\pm = \frac{q_{{}_\pm}}{\sqrt{\chi^2 + (z \mp z_0)^2}} \ ,
\end{align}
where $\chi^{2}=x^{2}+y^{2}$. The spatial leaf $\mathcal V$, over which we calculate the interaction Lagrangian, is the three-dimensional Euclidean space without two points at $z=\pm z_0$ on the $z$-axis: $\mathcal V=\mathbb R^3 - \{(0, 0, \pm z_0)\}$ and $\Omega = 4\pi$. The interaction Lagrangian in \eqref{eq:int_Lagrangian_potentials} has three sectors: two inner ones, each around either of the point singularities, and an outer one at infinity. The latter vanishes due to the fall-off behavior of the fields. To evaluate the surface integrals over the inner boundaries, we consider those boundaries to be two cylindrical pill-boxes $\mathcal P_\pm$ around the point singularities. The two surface integrals in interaction Lagrangian add up to
\begin{align} \label{eq:Coulomb_int_Lagrangian}
\mathcal L_{\text{int}} \ & = \  \frac{1}{8 \pi}\left(\oint_{\mathcal P_+} \boldsymbol \nabla(\mathcal{Z}_+\,  \mathcal{Z}_-) \cdot \mathrm{d} \boldsymbol S
+  \oint_{\mathcal P_-} \boldsymbol \nabla (\mathcal{Z}_+\,  \mathcal{Z}_-) \cdot \mathrm{d} \boldsymbol S \right) \nonumber \\
	& =  \ \left(q_{{}_+} q_{{}_-}\right)\frac{1}{d} \ ,
\end{align}
which is, as expected, proportional to the inverse of the distance $d=2z_0$ between the fields' singular points. The calculation here is a very much simpler version of the calculation that we perform for two magic fields in the following section, where $a_{{}_\pm}$ vanish and the space is single-sheeted. Therefore, we skip the details of the calculation here. Two alternative approaches for computation of Coulomb interaction from the action are presented in the appendices: one using the separating plane method in Appendix \ref{sec:app_SP_metod}, and the other using the bi-spherical coordinates in Appendix \ref{sec:app_Coulomb_bispherical}.

As for the interacting magic fields, we need to analyze the topology of the Riemann surface $\mathcal V$ of the superposition $\mathcal F_{\text{tot}}$. Recall that a single magic field's potential $\Ptl$ is continuous and single-valued on a two-sheeted extended space. In the case of interacting magic fields, each of the potentials\footnote{The sign indices $\pm$ of potential $\mathcal Z_\pm$ in this general consideration only refer to either of the interacting fields, but in the subsequent sections, they refer to upper/lower or right/left position of the field singularities, respectively.}, $\Ptl_{+}$ and $\Ptl_{-}$, can be in one of two branch configurations on a given sheet. Therefore, the superposed potential $\Ptl_{\text{tot}} = \Ptl_+ + \Ptl_-$ is continuous on a Riemann surface consisting of four sheets\footnote{The fact that the spacetime of two ring singularities needs to be four-sheeted (and, more generally, $2^N$-sheeted when there are $N$ such rings) is also noted in \cite{Tahvildar-Zadeh2019}.}. Figure \ref{fig:TopoInteractingMF} shows the topology of the entire Riemann surface of two interacting magic fields and a curve that initiates from one sheet, passes through the gates to all other sheets, and returns back to the beginning point.

\newcommand{\sx}{-5}
\newcommand{\sy}{-3}
\newcommand{\ssx}{5}
\newcommand{\ssy}{-3}
\newcommand{\spx}{0}
\newcommand{\spy}{-6}
\newcommand{\cpathone}{blue!40!gray!80!white}
\newcommand{\cpathtwo}{black}
\newcommand{\gblue}{gray!50!white}

\begin{figure}[h]
	\centering
	\scalebox{.8}{
		\begin{tikzpicture}
			\draw (-0.1,2.33) node[ellipse, color=red, minimum height=1.05cm,minimum width=2.6cm,draw,thick] {};
			\draw (-0.1,0.33) node[ellipse, color=blue, minimum height=1.05cm,minimum width=2.6cm,draw,thick] {};
			
			\node[align=center] at (-0.1,4.5) {%
				$\Ptl_{+}+\Ptl_{-}$\\
				$\Ptl_{+} \Ptl_{-}$};
            \node[align=center] at (-1.9,3.2) {$\mathcal{V}_{1}$};
			\draw[-, color=\gblue, dashed] (-2.33,3.66) -- (2.14,3.66) ;
			\draw[-, color=\gblue, dashed] (-2.33,3.66) -- (-2.33,-1) ;
			\draw[-, color=\gblue, dashed] (2.14,3.66) -- (2.14,-1) ;
			\draw[-, color=\gblue, dashed] (-2.33,-1) -- (2.14,-1) ;

			\draw (-0.1+\sx,2.33+\sy) node[ellipse, color=red, minimum height=1.05cm,minimum width=2.6cm,draw,thick] {};
			\draw (-0.1+\sx,0.33+\sy) node[ellipse, color=blue, minimum height=1.05cm,minimum width=2.6cm,draw,thick] {};
			
			\node[align=center] at (-0.1+\sx,4.5+\sy) {%
				$-\Ptl_{+}+\Ptl_{-}$\\
				$-\Ptl_{+} \Ptl_{-}$};
            \node[align=center] at (-1.9+\sx,3.2+\sy) {$\mathcal{V}_{2}$};
			\draw[-, color=\gblue, dashed] (-2.33+\sx,3.66+\sy) -- (2.14+\sx,3.66+\sy) ;
			\draw[-, color=\gblue, dashed] (-2.33+\sx,3.66+\sy) -- (-2.33+\sx,-1+\sy) ;
			\draw[-, color=\gblue, dashed] (2.14+\sx,3.66+\sy) -- (2.14+\sx,-1+\sy) ;
			\draw[-, color=\gblue, dashed] (-2.33+\sx,-1+\sy) -- (2.14+\sx,-1+\sy) ;

			\draw (-0.1+\ssx,2.33+\ssy) node[ellipse, color=red, minimum height=1.05cm,minimum width=2.6cm,draw,thick] {};
			\draw (-0.1+\ssx,0.33+\ssy) node[ellipse, color=blue, minimum height=1.05cm,minimum width=2.6cm,draw,thick] {};
			
			\node[align=center] at (-0.1+\ssx,4.5+\ssy) {%
				$\Ptl_{+}-\Ptl_{-}$\\
				$-\Ptl_{+} \Ptl_{-}$};
            \node[align=center] at (-1.9+\ssx,3.2+\ssy) {$\mathcal{V}_{4}$};
			\draw[-, color=\gblue, dashed] (-2.33+\ssx,3.66+\ssy) -- (2.14+\ssx,3.66+\ssy) ;
			\draw[-, color=\gblue, dashed] (-2.33+\ssx,3.66+\ssy) -- (-2.33+\ssx,-1+\ssy) ;
			\draw[-, color=\gblue, dashed] (2.14+\ssx,3.66+\ssy) -- (2.14+\ssx,-1+\ssy) ;
			\draw[-, color=\gblue, dashed] (-2.33+\ssx,-1+\ssy) -- (2.14+\ssx,-1+\ssy) ;

			\draw (-0.1+\spx,2.33+\spy) node[ellipse, color=red, minimum height=1.05cm,minimum width=2.6cm,draw,thick] {};
			\draw (-0.1+\spx,0.33+\spy) node[ellipse, color=blue, minimum height=1.05cm,minimum width=2.6cm,draw,thick] {};
			
			\draw[-, color=\gblue, dashed] (-2.33+\spx,3.66+\spy) -- (2.14+\spx,3.66+\spy) ;
			\draw[-, color=\gblue, dashed] (-2.33+\spx,3.66+\spy) -- (-2.33+\spx,-1+\spy) ;
			\draw[-, color=\gblue, dashed] (2.14+\spx,3.66+\spy) -- (2.14+\spx,-1+\spy) ;
			\draw[-, color=\gblue, dashed] (-2.33+\spx,-1+\spy) -- (2.14+\spx,-1+\spy) ;
			\node[align=center] at (-0.1+\spx,-2+\spy) {%
				$-\Ptl_{+}-\Ptl_{-}$\\
				$\Ptl_{+} \Ptl_{-}$};
            \node[align=center] at (-1.9+\spx,3.2+\spy) {$\mathcal{V}_{3}$};
			
			\draw[color=\cpathone,thick,dashed,->] (-0.1,2.33)..controls (-0.1,2.33-1) and (-0.1-1.3+1,1.03)..(-0.1-1.3,1.03)..controls (-0.1-1.3-0.5,1.03) and (-0.1+\sx,3.03+\sy+1)..(-0.1+\sx,2.33+\sy);
			\draw[color=\cpathtwo,thick,->] (-0.1+\sx,2.33+\sy) -- (-0.1+\sx,0.33+\sy);
			\draw[color=\cpathone,thick,dashed,->] (-0.1+\sx,0.33+\sy)..controls (-0.1+\sx,0.33+\sy-2.5) and (-0.1-1.8/2+\spx-2.5,0.33+\spy+1)..(-0.1-1.8/2+\spx,0.33+\spy+1)..controls (-0.1-1.8/2+\spx+1,0.33+\spy+1) and (-0.1+\spx,0.33+\spy+0.8).. (-0.1+\spx,0.33+\spy);
			\draw[color=\cpathtwo,thick,->](-0.1+\spx,0.33+\spy)..controls (-0.1+\spx,0.33+\spy-1) and (-0.1+\spx+1.8,0.23+\spy-1)..(-0.1+\spx+1.8,0.43+\spy);
			\draw[color=\cpathtwo,thick,<-](-0.1+\spx,2.33+\spy)..controls (-0.1+\spx,2.33+\spy+1) and (-0.1+\spx+1.8,2.23+\spy+1)..(-0.1+\spx+1.8,2.23+\spy)..controls (-0.1+\spx+1.8,2.23+\spy-1) and (-0.1+\spx+1.8,-0.66+\spy+1)..(-0.1+\spx+1.8,0.43+\spy);
			\draw[color=\cpathone,dashed,thick,<-](-0.1+\spx+1.8,2.23+\spy-0.86)..controls (-0.1+\spx+1.8-1,2.23+\spy-0.86) and (-0.1+\spx,2.33+\spy-1)..(-0.1+\spx,2.33+\spy);
			\draw[color=\cpathone,dashed,thick,-](-0.1+\spx+1.8,2.23+\spy-0.86)--(-0.1+\ssx,2.23+\spy-0.86);
			\draw[color=\cpathone,thick,dashed,<-](-0.1+\ssx,2.33+\ssy)..controls (-0.1+\ssx,2.33+\ssy+1) and (-0.1+\ssx+1.8,2.23+\ssy+1)..(-0.1+\ssx+1.8,2.23+\ssy)..controls (-0.1+\ssx+1.8,2.23+\ssy-1) and (-0.1+\ssx+1.8,-0.66+\ssy+1)..(-0.1+\ssx+1.8,-0.66+\ssy);
			\draw[color=\cpathone,thick,dashed,<-](-0.1+\ssx+1.8,-0.66+\ssy)..controls(-0.1+\ssx+1.8,-0.66+\ssy-1) and (-0.1+\ssx+1,2.23+\spy-0.86).. (-0.1+\ssx,2.23+\spy-0.86); 
			\draw[color=\cpathtwo,thick,<-] (-0.1+\ssx,0.33+\ssy) -- (-0.1+\ssx,2.33+\ssy);
			\draw[color=\cpathone,thick,dashed,<-](-0.1-2.5+\ssx,0.33+\ssy)..controls (-0.1-2.5+\ssx,0.33+\ssy-1) and (-0.1+\ssx,0.33+\ssy-1)..(-0.1+\ssx,0.33+\ssy);
			\draw[color=\cpathone,dashed,thick,->](-0.1-2.5+\ssx,0.33+\ssy)--(-0.1-2.5+\ssx,0.43-1);
			\draw[color=\cpathone,dashed,thick,->](-0.1-2.5+\ssx,0.43-1)..controls(-0.1-2.5+\ssx,-0.1+1.2) and (-0.1+1.2+1,0.33+0.86)..(-0.1+1.2,0.33+0.86);
			\draw[color=\cpathone,dashed,thick,->](-0.1+1.2,0.33+0.86)..controls (-0.1+1.2-.5,0.33+0.86) and (-0.1,0.33+1.0)..(-0.1,0.23);
			\draw[color=\cpathtwo,thick,->](-0.1,0.33)..controls (-0.1,0.33-1) and (-0.1+1.8,0.23-1)..(-0.1+1.8,0.43);
			\draw[color=\cpathtwo,thick,<-](-0.1,2.33)..controls (-0.1,2.33+1) and (-0.1+1.8,2.23+1)..(-0.1+1.8,2.23)..controls (-0.1+1.8,2.23-1) and (-0.1+1.8,-0.66+1)..(-0.1+1.8,0.43);
		
	\end{tikzpicture}}
	\caption{Topological structure of the Riemann surface of two interacting Magic Fields. Each gray dashed square is a distinct sheet. A curve (solid line) passes through the gates (disks) and goes from one sheet to another. Dashed lines demonstrate such passing. The branch of the potentials at each sheet is indicated by the relative sign with respect to the principal branch for both superposition and product.}
	\label{fig:TopoInteractingMF}
\end{figure}
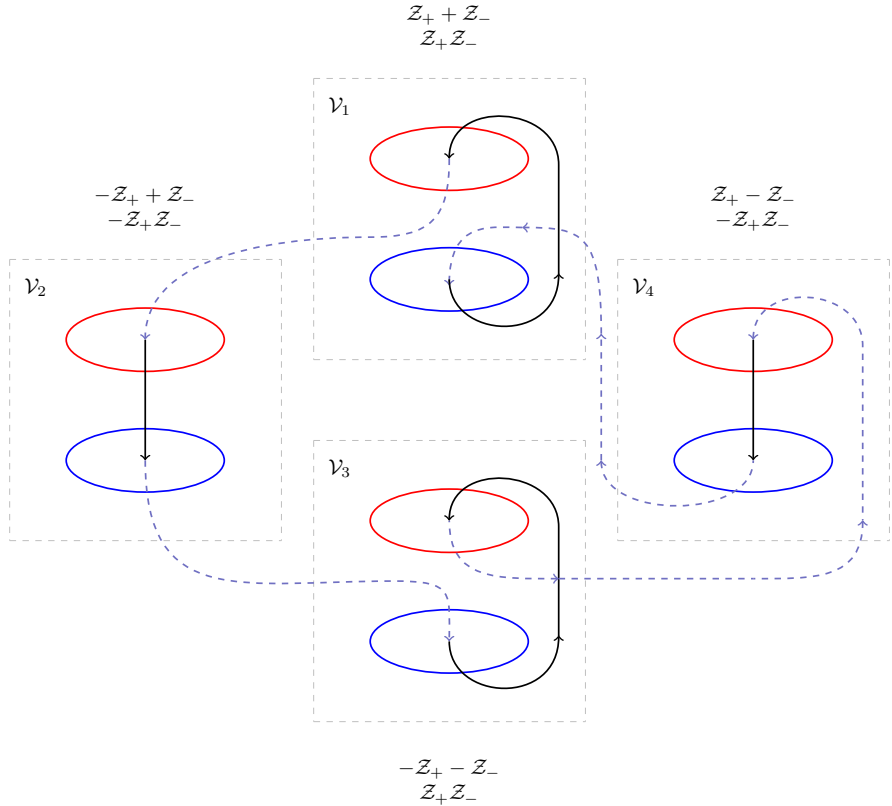

The interaction Lagrangian \eqref{eq:int_Lagrangian_potentials} consists of surface integrals 
over the boundaries $\partial\mathcal{V}$ of the four-sheeted Riemann surface $\mathcal V$.
However, the complex function $\mathcal{Z}_+ \mathcal{Z}_-$ has only two branches and has different overall signs in two pairs of equivalent sheets as written in Figure \ref{fig:TopoInteractingMF}. These signs cancel the sign change due to the change of orientation of the coordinate system, and therefore, the integrals have the same value on all sheets. Moreover, in this case, instead of tube-like surfaces that cover the ring singularities in both sheets, integrating over two closed infinitesimal surfaces $\mathcal P_\pm$ around each singularity in either of the sheets yields the same result. Taking the overall factor in \eqref{eq:int_Lagrangian} yields $4 \times 1/2\,\Omega = 1/8\pi$ into account, the interaction Lagrangian takes a similar form to the Coulomb case as
\begin{align} \label{eq:EMF_int_Lagrangian}
\mathcal L_{\text{int}} \ & = \ \frac{1}{8\pi} \left(\oint_{\mathcal P_+} \boldsymbol \nabla(\mathcal{Z}_+\,  \mathcal{Z}_-) \cdot \mathrm{d} \boldsymbol S
+  \oint_{\mathcal P_-} \boldsymbol \nabla (\mathcal{Z}_+\,  \mathcal{Z}_-) \cdot \mathrm{d} \boldsymbol S \right) \ .
\end{align}
We evaluate this in two different arrangements, with the ring singularities located in up-down and side-by-side configurations.

\subsection{Interaction of magic fields in up-down configuration}

In this section, we study the interaction of two electromagnetic magic fields in an up-down configuration. Consider two electromagnetic magic fields, with their ring singularities on $z=\pm z_0$ and oriented along the $z$-axis with potentials
\begin{equation}\label{eq:Up_down_potntials}
\mathcal{Z}_\pm = \frac{q_{{}_\pm}}{\sqrt{\chi^{2}+(z \mp z_0- i \,a_{{}_\pm})^{2}}} \ ,
\end{equation}
where the plus and minus signs refer to the upper and lower rings, which have, in general, different electric charges $q_{{}_\pm}$ and spin parameters $a_{{}_\pm}$. 

\begin{figure}[h]
    \centering
    \includegraphics[width=8cm]{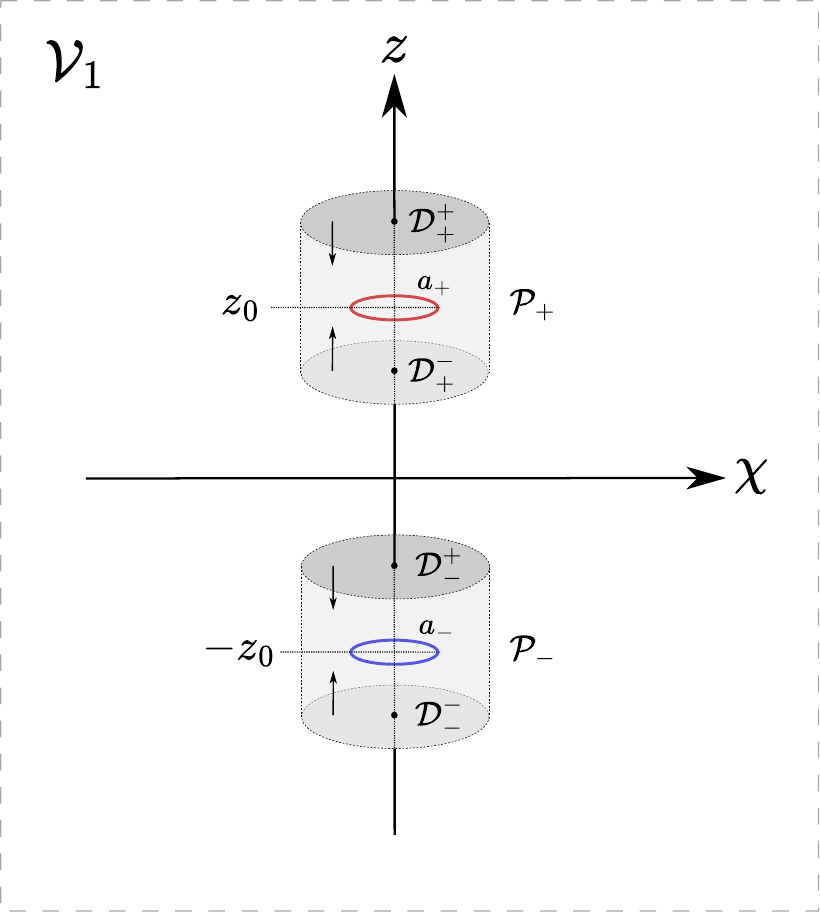}
    \caption{Up-down magic field configuration}
    \label{image_up_down_2}
\end{figure}

We evaluate the surface integrals in \eqref{eq:EMF_int_Lagrangian} on two cylindrical pillboxes $\mathcal{P}_{\pm}$, around the two ring singularities, as shown in Figure \ref{image_up_down_2}. The azimuthal symmetry of the problem in integrands and surfaces allows us to exploit the polar coordinates $(\chi,\phi,z)$ to expand the involving integrals as
\begin{align}\label{eq:IL_integrals1}
\oint_{\mathcal P_\pm} \boldsymbol \nabla(\mathcal{Z}_+\,  \mathcal{Z}_-) \cdot \mathrm{d} \boldsymbol S \
& =   \
- \int_{\mathcal D_{\pm}^+} \,
\frac{\partial}{\partial z}(\mathcal Z_+\, \mathcal Z_-) \,   d S
+ \int_{\mathcal D_{\pm}^-}
\frac{\partial}{\partial z}(\mathcal Z_+\, \mathcal Z_-) \, d S
\end{align}  
where $d S = \chi d\chi d\phi$ is the area element on upper $\mathcal D_+^+$ and lower $\mathcal D_+^-$ disk-like bases of the upper pillbox $\mathcal P_+$ (and similarly, bases $\mathcal D_-^\pm$ of the lower pillbox $\mathcal P_-$) in polar coordinates, while the vanishing parts of the integrals over the cylindrical sides of the pillboxes are eliminated. The signs of each term in \eqref{eq:IL_integrals1} are produced by the direction of the normals of the surfaces, which are toward inside the pillboxes.

The pillboxes $\mathcal P_\pm$ are infinitesimal surfaces, and we have to take the appropriate limits so that their heights tend to zero and the radius of each of them tends to the radius of the covered ring singularity from outside. The order of integration, adding up the contributions for each pillbox, and taking the limits are important in this computation. This is because the limiting process makes the phase of $\mathcal{Z}_+  \mathcal{Z}_-$ ambiguous during the integration. In other words, taking the limit before integration leads to incorrect results.

The integrals in the right-hand sides of equations \eqref{eq:IL_integrals1} over all four disks $\mathcal D_\pm^\pm$ are of the same form, and we can calculate them by using the primitive integral
\begin{align}\label{eq:primitive_func}
F(\chi,z)\ := &\ \int \frac{\partial}{\partial z}\left(\frac1{\sqrt{\chi^2+A_+(z)}\,\sqrt{\chi^2+A_-(z)}}\right)\chi\,d\chi 
\nonumber\\
= & \ \frac1{2\,(A_+-A_-)}\left(-A_+'\,\frac{\sqrt{\chi^2+A_-}}{\sqrt{\chi^2+A_+}} + A_-'\, \frac{\sqrt{\chi^2+A_+}}{\sqrt{\chi^2+A_-}}\right) + c
\end{align}
for $A_\pm \in \mathbb{C}$ and $\chi \in \mathbb{R}$ and substitutions
\begin{align}\label{eq:A_pm}
    A_\pm(z) \ &= \ (z \mp z_0-i\, a_{{}_\pm})^2 \quad ,\quad A_\pm'(z) \ = \ 2\,(z \mp z_0-i\, a_{{}_\pm}) \ . 
\end{align}
Using this relation and substitutions and specifying the proper limits for infinitesimal disks $\mathcal D_\pm^\pm$, we obtain the right-hand side of \eqref{eq:IL_integrals1} as
\begin{align}\label{eq:IL_integrals2}
\oint_{\mathcal P_\pm} \boldsymbol \nabla(\mathcal{Z}_+\,  \mathcal{Z}_-) \cdot \mathrm{d} \boldsymbol S \
& =   \ 2\pi\,q_{{}_+}q_{{}_-} \times \lim\limits_{a\to |a_{{}_\pm}|} \left[
-\lim\limits_{z\to \pm z_0^+} F(\chi,z)\Big|_{\chi=0}^{\chi=a} + \lim\limits_{z\to \pm z_0^-} F(\chi,z) \Big|_{\chi=0}^{\chi=a} \right] \ ,
\end{align}
where the factor of $2\pi$ is the result of integration over azimuthal angular coordinate $\phi$. Note that the order of the limits is crucial in this calculation, as we will see in the following.

\begin{figure}[htp]
    \centering
    \includegraphics[width=8cm]{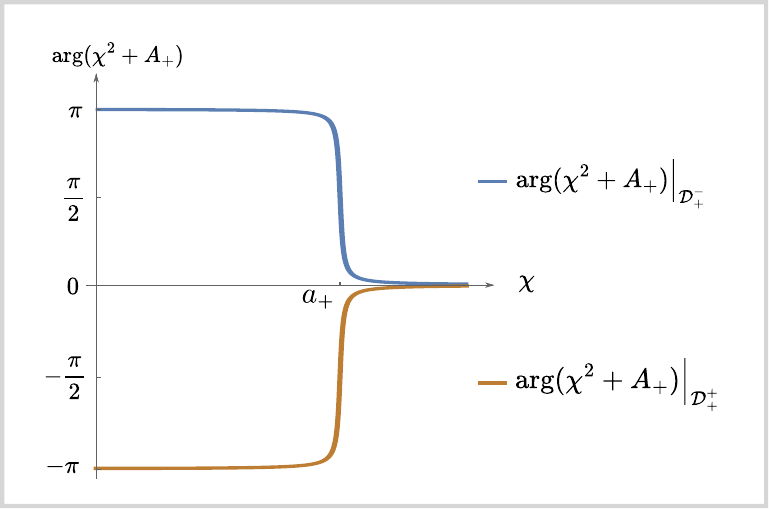}
    \caption{Argument of the radicant of $\sqrt{\chi^2+A_+}$, with $A_+ = (\epsilon - i\,a_{{}_+})^2$ for infinitesimal $\epsilon \to 0^\pm$ over the disk $\mathcal D_+^\pm$.}
    \label{fig:arg}
\end{figure}

In order to proceed further and find the values of the terms in \eqref{eq:IL_integrals2}, we need to analyze the square root functions in \eqref{eq:primitive_func} with \eqref{eq:A_pm}, and consistently choose the branches for them. To be more specific, consider the square root function $\sqrt{\chi^2+A_+}$, with $A_+$ in \eqref{eq:A_pm}, that is in fact the inverse of the potential $\mathcal Z_+$. This function vanishes on the upper ring singularity $(\chi = |a_{{}_+}|, z =z_0)$. Around this ring, i.e. over the upper/lower base $\mathcal D_+^\pm$ of the upper pillbox $\mathcal P_+$, we need to analyze the limits
\begin{align}\label{eq:squre_root_anlyisis1}
   \sqrt{\chi^2 + A_+} \,\Big|_{\mathcal D_+^\pm} \ = \ \lim\limits_{z\to z_0^{\pm}}\sqrt{\chi^2 + (z - z_0 - i\, a_{{}_+})^2} \ = \ \lim\limits_{\epsilon\to 0^{\pm}}\sqrt{\chi^2 + (\epsilon - i \,a_{{}_+})^2} \ .
\end{align}
The argument of the radicant of this function is in the intervals
\begin{align}
    \arg({\chi^2+A_+}) \in 
    \begin{cases}
        (-\pi,0) & \text{over upper disk} \ \mathcal D_+^+ \ , \\
        (0,\pi) & \text{over lower disk} \ \mathcal D_+^- \ ,
    \end{cases}
\end{align}
and its behaviour is depicted in figure \ref{fig:arg}.

Using the principal branch cut at $\arg(\alpha)=\pm\pi$, i.e. the negative real axis, we can compute the contributions from each limit in \eqref{eq:IL_integrals2}. In particular, at the upper limit of integrals at $\chi = a > |a_{{}_+}|$, we have
\begin{align}
    \lim\limits_{\epsilon\to 0^{\pm}}\sqrt{\chi^2 +(\epsilon - i \,a_{{}_+})^2}\Big|_{\chi=a} \ = \ \sqrt{a^2-a_{{}_+}^2} \ > \ 0 \ ,
\end{align}
regardless of the disk. Since other ingredients of the function $F(\chi,z)$ have the same finite value on $\mathcal D_+^\pm$, the two terms in \eqref{eq:IL_integrals2}, due to their different signs, cancel out at this limit. Therefore, the limit $a\to a_{{}_+}$ yields no divergence but vanishes smoothly. 

On the other hand, at the lower limit $\chi = 0$, we have
\begin{align} \label{eq:Up-down_lim_sqrt}
    \lim\limits_{\epsilon\to 0^{\pm}}\sqrt{\chi^2 + (\epsilon - i \,a_{{}_+})^2} \, \Big|_{\chi=0} \ = \ \mp\, i \,a_{{}_+} \ .
\end{align}
Therefore, at this limit, we can explicitly derive
\begin{align}\label{eq:F+}
    &\lim\limits_{z\to z_0^+} F(\chi , z) \Big|_{\chi = 0}
    \nonumber\\
    & \ = \
    \lim\limits_{\epsilon\to 0^+} \frac1{(\epsilon-i\,a_{{}_+})^2 - (2\,z_0-i\,a_{{}_-})^2} 
    \left[
    -\,(\epsilon - i\,a_{{}_+}) \, \frac{\sqrt{(2\,z_0- i\,a_{{}_-})^2}}{\sqrt{(\epsilon - i\,a_{{}_+})^2}} 
    + (2\,z_0 - i\, a_{{}_-}) \, \frac{\sqrt{(\epsilon - i \, a_{{}_+})^2}}{\sqrt{(2\,z_0 - i\, a_{{}_-})^2}}
    \right]
    \nonumber\\
    & \ = \
    \frac1{(i\,a_{{}_+})^2 - (2\,z_0-i\,a_{{}_-})^2} \left[
    - \sqrt{(2\,z_0- i\,a_{{}_-})^2}
    + (2\,z_0 - i\, a_{{}_-}) \, \frac{ - i \, a_{{}_+}}{\sqrt{(2\,z_0 - i\, a_{{}_-})^2}}
    \right]
    \nonumber\\
    &\ = \ \frac1{2\,z_0 - i \, (a_{{}_+} + a_{{}_-})} \ ,
\end{align}
where for the last equal, we have used $z_0>0$, and thus $\sqrt{(2\,z_0 - i\,a_{{}_-})^2} = 2\,z_0 - i\,a_{{}_-}$. 
Now due to the signs of the terms in \eqref{eq:IL_integrals2} and additional signs from \eqref{eq:Up-down_lim_sqrt}, the contributions of the lower and upper disks $\mathcal D_+^\pm$ are equal, and therefore for the upper pillbox $\mathcal P_+$, we have
\begin{align}\label{eq:Upper_integral}
     \oint_{\mathcal P_+} \boldsymbol \nabla(\mathcal{Z}_+\,  \mathcal{Z}_-) \cdot \mathrm{d} \boldsymbol S 
    \ & = \ 
    4\pi\,q_{{}_+}q_{{}_-} \times \lim\limits_{z\to z_0^+} F(\chi^2 , z) \Big|_{\chi = 0}
    \nonumber\\
    \ & = \ \frac{4\pi\,q_{{}_+}q_{{}_-}}{2\,z_0 - i \, (a_{{}_+} + a_{{}_-})} \ ,
\end{align}

Similar analysis can be carried out for the other singularity-generating square root function $\sqrt{\chi^2+A_-}$ on the surfaces around the lower singularity at $(\chi= |a_{{}_-}| ,z=-z_0)$. We can employ the transformation ($a_{{}_+} \leftrightarrow a_{{}_-}$ and $z_0 \leftrightarrow -z_0$), relating the upper and lower potential functions $\mathcal Z_\pm$ in \eqref{eq:Up_down_potntials} and also in \eqref{eq:A_pm}, to transform the result found in \eqref{eq:Upper_integral} for the upper pillbox $\mathcal P_+$ to the result of integration on the lower pillbox $\mathcal P_-$. To do so, we must apply this transformation before simplifying the square roots. In particular in transforming \eqref{eq:F+}, we encounter $\sqrt{(-2\,z_0 - i\,a_{{}_+})^2} = 2\,z_0 + i\, a_{{}_+} $, which produces an extra overall minus sign. Performing this computation yields
\begin{align}
     \oint_{\mathcal P_-} \boldsymbol \nabla(\mathcal{Z}_+\,  \mathcal{Z}_-) \cdot \mathrm{d} \boldsymbol S 
    \ & = \ 
    4\pi\,q_{{}_+}q_{{}_-} \times \lim\limits_{z\to -z_0^+} F(\chi^2 , z) \Big|_{\chi = 0}
    \nonumber\\
    \ & = \ \frac{4\pi\,q_{{}_+}q_{{}_-}}{2\,z_0 + i \, (a_{{}_+} + a_{{}_-})} \ ,
\end{align}
which is the transformed \eqref{eq:Upper_integral} with an extra overall minus sign.

Finally, the total interaction Lagrangian \eqref{eq:EMF_int_Lagrangian} for two magic fields in up-down positions is obtained as
\begin{align} \label{eq:EMF_UD_int_Lagrangian2a}
\mathcal L_{\text{int}} \ & = \  \frac{1}{8\pi} \left( \frac{4\pi\,q_{{}_+}q_{{}_-}}{2\,z_0 - i \, (a_{{}_+} + a_{{}_-})} + \frac{4\pi\,q_{{}_+}q_{{}_-}}{2\,z_0 + i \, (a_{{}_+} + a_{{}_-})} \right)
\nonumber\\
\ & = \ (q_{{}_+}q_{{}_-})\,\frac{d}{d^2 + (a_{{}_+} + a_{{}_-})^2}
\end{align}
where $d=2z_0$ is the distance between the centers of the singularity rings.
Notice that the total interaction Lagrangian is real and, in the spin-zero limit, reproduces the Coulomb interaction. Since we haven't restricted the signs of $a_{{}_\pm}$ throughout the calculation, they can represent both aligned and anti-aligned spins.

\subsection{Interaction of magic fields in side-by-side configuration}

\begin{figure}[ht]
    \centering
    \includegraphics[width=10cm]{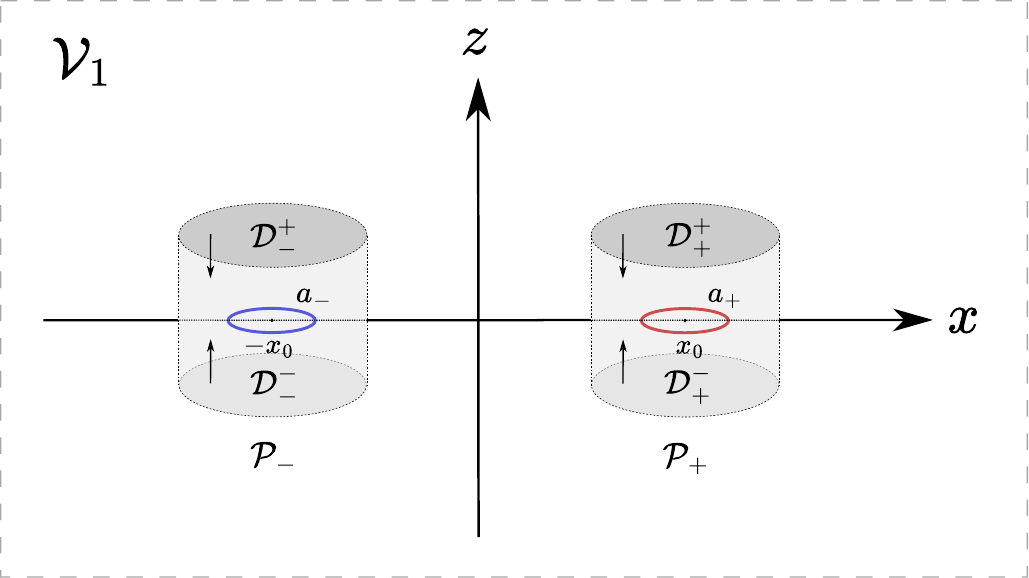}
    \caption{Side-by-side magic field configuration}
    \label{fig:image_side}
\end{figure}

In this section, we study the interaction of two electromagnetic magic fields in a side-by-side configuration. Consider two magic fields whose ring singularities are located on a plane with their centers at $x=\pm x_0$ separated by a distance $d=2x_0$, see figure \ref{fig:image_side}.

Like the previous section, the magic fields have, in general, different spin parameters, i.e. ring radii $a_{{}_+} \! \neq a_{{}_-}$. In Cartesian coordinates $(x,y,z)$, complex scalar potentials for magic fields are 
\begin{align} \label{eq:side_by_side_potntials}
 \Z\!{}_{{}_\pm} = \frac{q_{{}_\pm}}{\sqrt{(x \mp x_0)^{2} + y^2 +(z - i \,a_{{}_\pm})^{2}}} \ ,
\end{align}
The topology of the superposition field is the same as what we analyzed before and depicted in Figure \ref{fig:TopoInteractingMF}. The only difference is that the ring singularities are positioned side by side in each sheet. The interaction Lagrangian, therefore, has the same form as \eqref{eq:EMF_int_Lagrangian}
with $\mathcal P_\pm$ being the pair of left and right infinitesimal pillboxes in either of the sheets. This configuration has no azimuthal symmetry, and we need to introduce an adapted coordinate system to evaluate the integrals. However, we can expand the involving integrals in Cartesian coordinates over the bases of the pill-boxes as
\begin{align}\label{eq:IL_SS_integrals1}
\oint_{\mathcal P_\pm} \boldsymbol \nabla(\Z\!{}_{{}_+}  \Z\!{}_{{}_-}) \cdot \ud \boldsymbol S \
& =   \
- \int_{\mathcal D_{\pm}^+} \,
\frac{\partial}{\partial z}(\Z\!{}_{{}_+}  \Z\!{}_{{}_-}) \,   \ud S
+ \int_{\mathcal D_{\pm}^-}
\frac{\partial}{\partial z}(\Z\!{}_{{}_+}  \Z \!{}_{{}_-}) \, \ud S
\end{align}  
where $\ud S = \ud x \ud y$ is the area element on upper $\mathcal D_+^+$ and lower $\mathcal D_+^-$ disk-like bases of the right pillbox $\mathcal P_+$ (and similarly, bases $\mathcal D_-^\pm$ of the left pillbox $\mathcal P_-$) in Cartesian coordinates, while the vanishing parts of the integrals over the cylindrical sides of the pillboxes are eliminated. The signs of each term in \eqref{eq:IL_SS_integrals1} are produced by the direction of the normals of the surfaces, which are toward inside the pillboxes. Each of the integrals has the form
\begin{align}\label{eq:primitive_func_SS}
G(x,y,z)\ & := \ 
\int \frac{\partial}{\partial z}\left(\frac1{\sqrt{(x - x_0)^2 + y^2 - \hat a_{{}_+}^2}\,\sqrt{(x + x_0)^2 + y^2 - \hat a_{{}_-}^2}}\right)\!\ud x\ud y
\end{align}
where $\hat a_{{}_\pm}$ are complex functions of $z$ as
\begin{align}\label{eq:B_pm}
          \hat a_{{}_\pm} \ = \ a_{{}_\pm} - i\,z \ . 
\end{align}

The attempt to work out the integral \eqref{eq:primitive_func_SS} in Cartesian coordinates leads to elliptic functions, and the analysis gets even more complicated. We adopt another approach, which consists of a coordinate transformation. The symmetries of the configuration would fit the bi-cylindrical coordinates if $\hat a_{{}_\pm}$ were real. Indeed, this is the case in the limit $z\to 0$, but we need first to integrate and then take this limit from different directions, i.e. above and below the $z=0$ plane, to find the correct contributions of the disks above and below the singular rings. A reliable approach would be to expand the integrand around $z=0$, analyze the integrals, and take the limits afterward. Following this approach, we will have over each disk $\mathcal D_\pm^\pm$ an expansion like
\begin{align}\label{eq:primitive_func_SS_3a}
    G(x,y,z) \ = \ G^0 (x,y) + \mathcal O(z) \ ,
\end{align}
where $G^0(x,y)$ are real functions. This allows us to analyze the branches of the square roots and determine the signs before integration, which requires a coordinate transformation.

To determine the correct signs at each disk, we should notice that inside each of the rings, 
the real part of the corresponding radiant for $z \ll 1$ is negative
\begin{align}
    (x \mp x_0)^2 + y^2 - a_{{}_\pm}^2 + z^2 < 0 \ .
\end{align}
This leads to different signs for that square root on the
upper and lower disks. On the contrary, the real part is positive for the points outside the ring, leading to the same sign for the square root on the above and below disks. As a result, considering their different signs in \eqref{eq:IL_SS_integrals1}, the integrals over the disks below and above the same ring singularity, at the integration limit inside the ring, will add up, while at the other limit outside the ring, i.e. the outer boundary of the disk, they will cancel out as $z$ goes to zero. Done with the branch cut analysis, we can further evaluate the real integral $G^0$ by transforming Cartesian coordinates to bi-polar coordinates.

In the following, we implement a more elegant approach without performing the expansion \eqref{eq:primitive_func_SS_3a}, keeping the complex nature of the integrand and applying a complex coordinate transformation.

Performing the derivative with respect to $z$ in \eqref{eq:primitive_func_SS}, by the Leibniz rule we get two terms
\begin{align}\label{eq:primitive_func_SS_2}
    G(x,y,z) \  = \ G_+(x,y,z) + G_-(x,y,z)
\end{align}
where $G_\pm(x,y,z)$ are
\begin{align}\label{eq:primitive_func_SS_3}
    G_\pm(x,y,z) \ = \  \int \frac{i\,\hat a_{{}_\pm}\,\ud x\ud y}{\left((x \mp x_0)^2 + y^2 - \hat a_{{}_\pm}^2 \right)^{3/2}\,\sqrt{(x \pm x_0)^2 + y^2 - \hat a_{{}_\mp}^2}} \ .
\end{align}

We assume that the singular rings do not overlap. It is equivalent to the condition
\begin{align}
    |a_{{}_+}|+|a_{{}_-}| < 2\, x_0 \, ,
\label{eq:distance_asumption}
\end{align}
where $2\, x_0 = d$ is the distance between centers of the singular rings. 

\begin{figure}[htp]
    \centering
    \includegraphics[width=12cm]{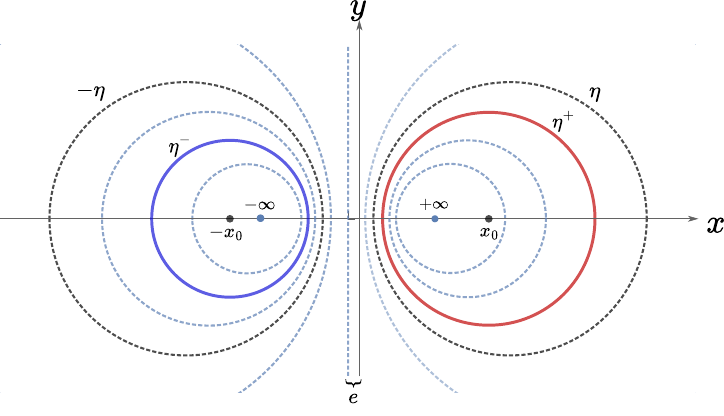}
    \caption{Bi-polar coordinates on $z=0$ plane: two sets of circles as the level sets of positive and negative values of $\eta$, including the ring singularities at $\eta^+$ and $\eta^-$ in red and blue, and the limits of the integral at $\pm \eta$ and $\pm \infty$}
    \label{fig:bpcoords}
\end{figure}

It is convenient to compute the integrals in bipolar coordinates. We consider two-dimensional coordinate transformation between $(x,y)$ and $(\eta, \sigma)$ in the form
\begin{subequations}
\begin{align}\label{eq:bipolar_tr}
x & \ = \ \hat f\, \frac{\sinh \eta}{\cosh \eta-\cos \sigma} - \hat e
\ ,
\\
y & \ = \  \hat f\, \frac{\sin \sigma}{\cosh \eta-\cos \sigma}
\ ,
\end{align} 
\end{subequations}
where $f$ and $e$ are parameters of the transformation which depend on $z$. The area element reads
\begin{equation}
\ud x\ud y \ = \ \frac{\hat f^{2}}{(\cosh \eta-\cos \sigma )^{2}} \, \ud\rho\ud\sigma \ .
\end{equation}
We choose the parameters of the transformation $\hat f$ and $\hat e$ to be
\begin{align}\label{eq:tr_parms}
\hat f \ &= \ \frac{1}{2d}\sqrt{d^4 - 2 d^2 \, (\hat a_{{}_+}^2 + \hat a_{{}_-}^2) + (\hat a_{{}_+}^2 - \hat a_{{}_-}^2)^2}
\ ,
\\
\hat e \ &= \ 
\frac{1}{2d}\left(\hat a_{{}_+}^2 - \hat a_{{}_-}^2\right)
\, .
\end{align} 
Since incorporating $a_{{}_\pm}$, these parameters are complex functions of $z$ and are chosen so that on the $z=0$ plane, they are real, and the singular rings of the magic fields are described by $\eta^\pm = \mathrm{const}$, see Figure \ref{fig:bpcoords}. 

The consequence of choosing the transformation parameters to be complex as \eqref{eq:tr_parms} is that the new coordinates $(\eta,\sigma)$ are generally complex.
In fact, the real $(x,y)$ plane is mapped into a subset of a two-dimensional complex space determined by a set of constraints. The transformation can also be written in the form
\begin{align}
    x + i\,y \ = \ \hat f\, \coth \frac{\eta - i\,\sigma}2 - \hat e \ .
\end{align}
The $\eta$-constant loci are circles described by
\begin{align}
    \left(x - c_\eta \right)^2 + y^2 \ = \ {r_\eta}^2 \ ,
\end{align}
with their centers and radii given by 
\begin{align}
    c_\eta \ = \ \hat f\, \coth\eta + \hat e \quad , \quad r_\eta \ = \ \frac{\hat f}{\sinh\eta} \ .
\end{align}
We consider the surfaces $\hat D^\pm_\pm$ consisting of the circles of constant $\eta$ that intersect with real $(x,y)$-plane on the $x$-axis, while the angle $\sigma$ spans each circle.

Using the transformation \eqref{eq:bipolar_tr}, we have
\begin{align}
(x - x_0)^2 + y^2 - \hat a_{{}_+}^2 \ & = \  \frac{2\hat f}{\cosh \eta - \cos \sigma} \left(  \hat f \cosh \eta - \hat b_{{}_+} \, \sinh \eta\right)
\\
(x + x_0)^2 + y^2 - \hat a_{{}_-}^2 \ & = \ \frac{2 \hat f}{\cosh \eta - \cos \sigma} \left( \hat f \cosh \eta + \hat b_{{}_-} \, \sinh \eta\right)
\end{align}
with
\begin{align}
    \hat b_{{}_\pm} \ := \ \sqrt{\hat f^2 + \hat a_{{}_\pm}^2 } \ = \ x_0 \pm \hat e
    \ .
\end{align}
As a result, the integrals $G_\pm$ in \eqref{eq:primitive_func_SS_3} transform into
\begin{align}
    G_\pm(\eta,\sigma,z) \ & = \ \frac14 \int \frac{ i\,\hat a_{{}_\pm} \,\hat f^2\, \ud\eta\ud\sigma}{\left(\hat f^2\,\cosh\eta \mp \hat f\, \hat b_{{}_\pm} \,\sinh\eta\right)^{3/2}\,\sqrt{\hat f^2\,\cosh\eta \pm \hat f\, \hat b_{{}_\mp} \,\sinh\eta}}
    \nonumber \\
    & = \ \frac{\pm\, i\,\pi\, \hat a_{{}_\pm}}{\hat f \, (\hat b_{{}_+} + \hat b_{{}_-})}\frac{\sqrt{\hat f^2 \cosh\eta \pm \hat f\,\hat b_{{}_\mp} \, \sinh\eta}}{\sqrt{\hat f^2 \cosh \eta \mp \hat f\,\hat b_{{}_\pm} \, \sinh\eta }}
\end{align}

The integral on an arbitrary $z$-plane is to be taken on a complex path for the coordinates $(\eta, \sigma)$, constrained by the reality condition of $(x, y)$, from a finite value associated with the outer circle, up to the infinity associated with the focal point inside the ring.

\begin{align}
    \lim_{z \to 0^\pm} \lim_{\eta \to +\infty} G_+ \ & = \ 
    \frac{i \pi a_{{}_+}  }{2\,f\,x_0}\times\frac{\sqrt{ f + b_{{}_+}} \sqrt{f + b_{{}_-}}}{\mp i\,a_{{}_+}}
    \nonumber \\ & = \ 
    \frac{\mp \pi  }{2\,f\,x_0}\,\sqrt{ (f+x_0)^2 - e^2}  
\end{align}

\begin{align}
    \lim_{z \to 0^\pm} \lim_{\eta \to +\infty} G_- \ & = \ 
    \frac{-i \pi a_{{}_-}  }{2\,f\,x_0}\times\frac{\mp i\,a_{{}_+}}{\sqrt{ f + b_{{}_+}} \sqrt{f + b_{{}_-}}}
    \nonumber \\ & = \ 
    \frac{ \mp \pi\,  a_{{}_+} a_{{}_-}  }{2\,f\,x_0}\times\frac{1}{\sqrt{ (f+x_0)^2 - e^2}}
\end{align}
where the unhatted parameters are the values of z-dependant functions on  $z=0$ plane, which makes all of them to be real. In particular, $f$ and $e$ are
\begin{align}
f \ & := \ \hat f\,\big|_{z=0} \ =  \ \frac{1}{2d}\sqrt{\left(d^2 - (a_{{}_+} - a_{{}_-})^2\right)\left(d^2 -  (a_{{}_+} + a_{{}_-})^2\right)}
\\
e \ & := \ \hat e\,\big|_{z=0} \ =  \ \frac1{2d}\left(a_{{}_+}^2 - a_{{}_-}^2 \right)
\end{align}

Combining these results, we obtain the interaction Lagrangian for two magic fields in a side-by-side position as
\begin{align} \label{eq:EMF_UD_int_Lagrangian2}
\mathcal L_{\text{int}} \ & = \  \frac{1}{8\pi} \left(4\pi\,q_{{}_+}q_{{}_-}  \right) \frac{d\,(d+2f)-(a_{{}_+}-a_{{}_-})^2}{d\,f\,\sqrt{(d+2 f)^2 - 4 \,e^2}}
\nonumber\\
\ & = \left(q_{{}_+}q_{{}_-} \right) \frac{d\,(d+2f)-(a_{{}_+}-a_{{}_-})^2}{2\,d\,f\,\sqrt{(d+2 f)^2 - 4 \,e^2}}
\end{align}
where $d=2\,x_0$ is the distance between the centers of the singular rings.

\subsection{Interaction potential of two magic fields}

The interaction potential for two magic fields is obtained from $U = \mathcal L_{\text{int}}$. Here, we analyze the interaction potential in two configurations that we calculated in previous sections. For two magic fields in an up-down configuration, the interaction potential is obtained from \eqref{eq:EMF_UD_int_Lagrangian2a} as
\begin{align} \label{eq:EMF_UD_int_potential2a}
U_{\text{\tiny UD}} \ & = \  
 (q_{{}_+}q_{{}_-})\,\frac{d}{d^2 + (a_{{}_+} + a_{{}_-})^2} \ .
\end{align}
The form of the interaction potential \eqref{eq:EMF_UD_int_potential2a} for the up-down configuration was first stated in \cite[Eq. (2)]{2012mgm..conf.2041R}, based on heuristic calculations based on a single sheet background that did not take into account the topology of the analytic extension of the magic field. 

The potential $U_{\text{\tiny UD}}$ vanishes in the limit $d \to 0$ and has extrema at $d=\pm|a_{{}_+} + a_{{}_-}|$ with 
\begin{align}
    U_{\text{\tiny UD}}^{\text{e}} = \frac{q_{{}_+}q_{{}_-}}{2\,|a_{{}_+} + a_{{}_-}|}\ .
\end{align}

 In contrast to the interaction of point particles, the interaction potential contains potential wells that allow for equilibrium points and oscillations.
For long-range interactions $a_{{}_\pm}<<d$, the leading terms are
    \begin{equation}
    U_{\text{\tiny UD}} \approx \left(q_{{}_+}q_{{}_-} \right) \left[\,\frac{1}{d} - \frac{(a_{{}_+}+a_{{}_-})^2}{ d^3}+ \mathcal O\left(\frac{1}{d^5}\right) \right]
    \, .
    \end{equation}
    In particular, we recover the Coulomb interaction in the spin-zero limit $ a_{{}_\pm} \!\to\! 0$ .
 As special cases, we consider the spins of interacting magic fields being either aligned equal $a_{{}_\pm} = a$ as well as anti-aligned equal $a_{{}_\pm} = \pm a$ in up-down configuration. 
In the aligned case, we obtain
\begin{align}
\label{eq:EMF_UD_int_Lagrangian2-2-1}
U_{\text{\tiny UD}}\Big|_{a_{{}_\pm} = a} \ & = \ 
 \left(q_{{}_+}q_{{}_-} \right)\frac{d}{d^2 + 4\, a^2} \ ,
\end{align}
whereas in the anti-aligned case, we have
\begin{align} \label{eq:EMF_UD_int_Lagrangian2-2}
U_{\text{\tiny UD}}\Big|_{a_{{}_\pm} = \pm a} \ & = \ \left(q_{{}_+}q_{{}_-} \right)\frac{1}d \ .
\end{align}

\begin{figure}[h]
	\centering
    \begin{subfigure}[b]{0.49\textwidth}
		\includegraphics[width=.95\linewidth]{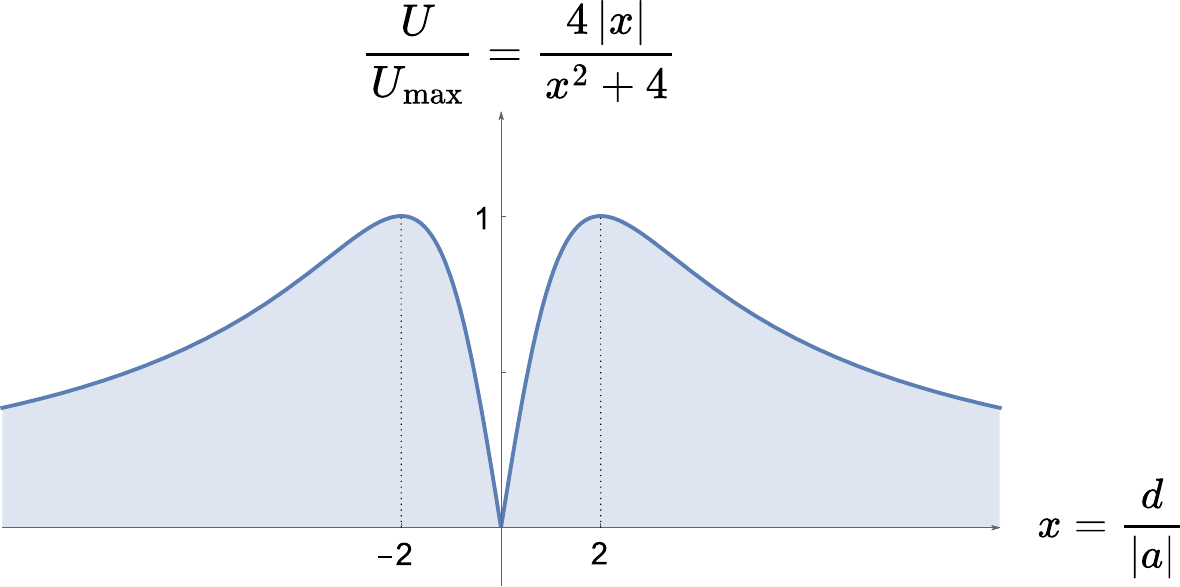}
		\caption{
        Up-down, $q_{1}q_{2}>0$ case: \\[5pt]
            $U=\dfrac{q_1q_2|d|}{d^2+4a^2} >0  $
        }	\label{fig:PotentialPlus}
    \vspace{.5cm}
    \end{subfigure}
    \begin{subfigure}[b]{0.49\textwidth}
		\includegraphics[width=.95\linewidth]{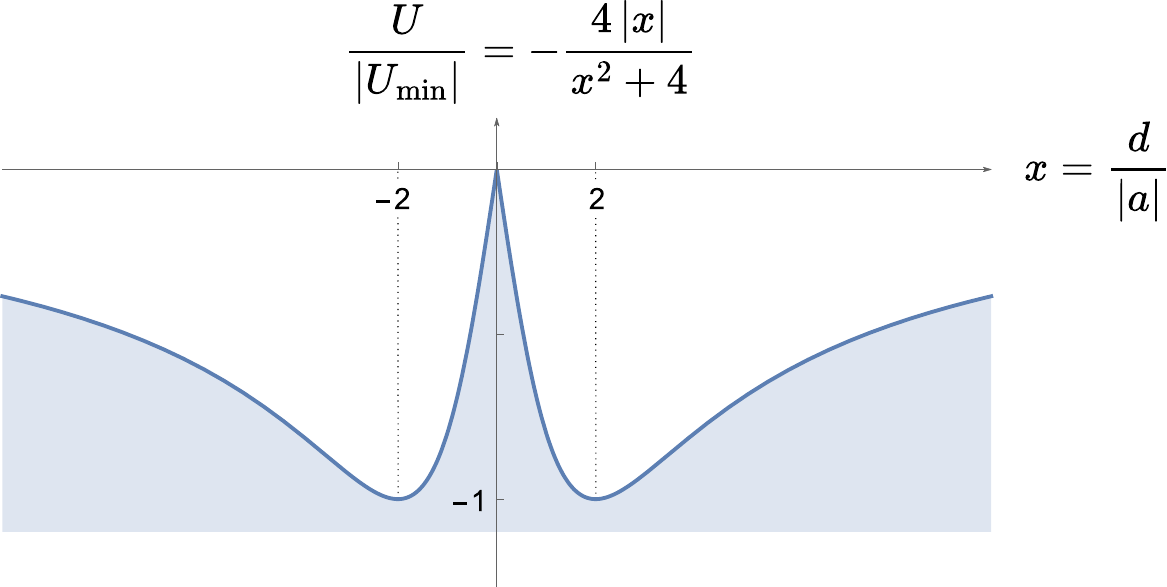}
		\caption{Up-down, $q_{1}q_{2}<0$ case: \\[5pt]
  $U=\dfrac{q_1q_2|d|}{d^2+4a^2} <0 $}
    \label{fig:PotentialMinus}
    \vspace{.5cm}
    \end{subfigure}
    
    \begin{subfigure}[b]{0.49\textwidth}
		\includegraphics[width=.95\linewidth]{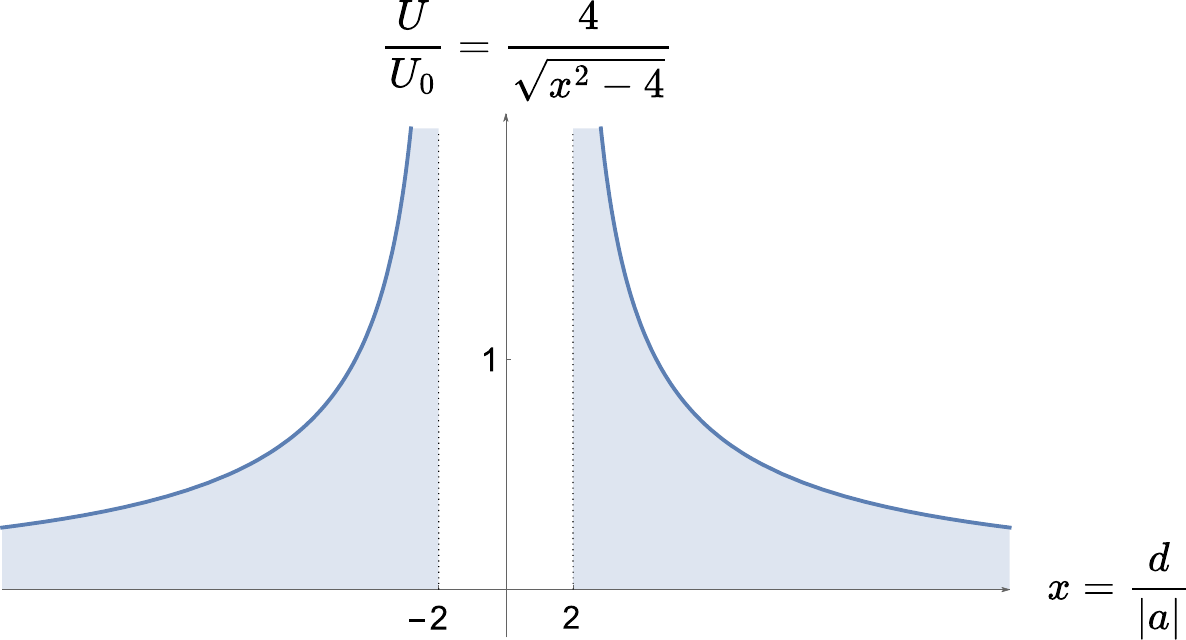}
		\caption{Side-by-side, $q_{1}q_{2}>0$ case: \\[5pt]
  $U=\dfrac{q_1q_2}{\sqrt{d^2-4a^2}} >0 $}
	\label{fig:PotentialPlus_ss}
    \end{subfigure}
    \begin{subfigure}[b]{0.49\textwidth}
		\includegraphics[width=.95\linewidth]{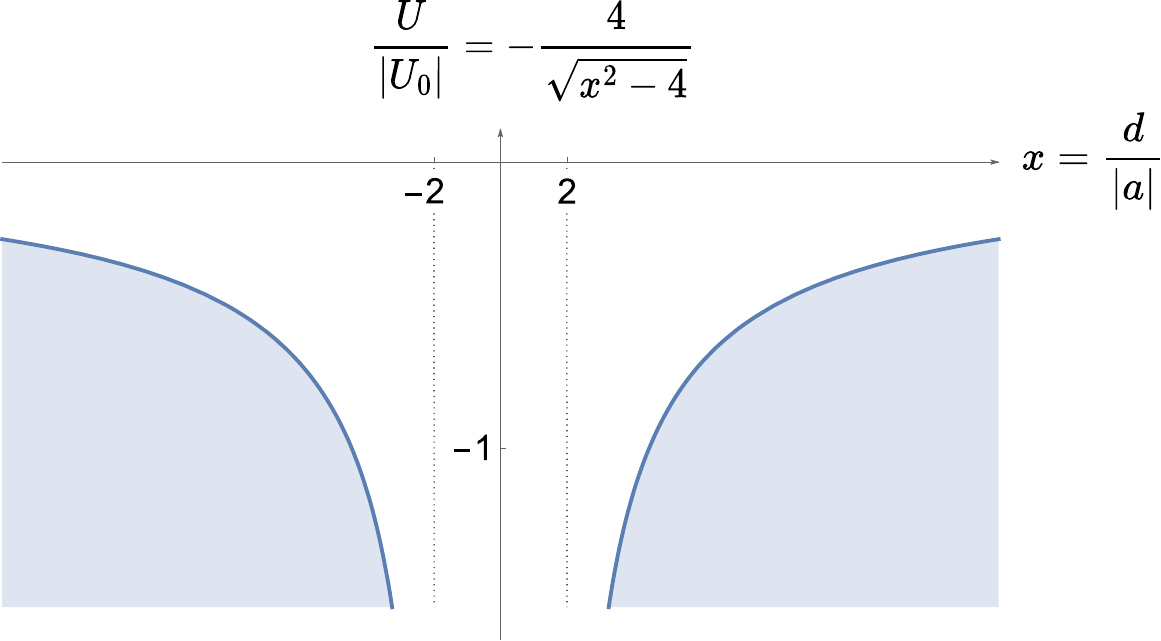}
		\caption{Side-by-side, $q_{1}q_{2}<0$ case: \\[5pt]
  $U=\dfrac{q_1q_2}{\sqrt{d^2-4a^2}} <0 $}
	\label{fig:PotentialMinus_ss}
    \end{subfigure}
    \caption{The interaction potential $U=\mathcal{L}_{\text{int}}$ of two magic fields in up-down and side-by-side configurations for $q_{1}\,q_{2}>0$ and  $q_{1}\,q_{2}<0$ cases, where the spin parameters are the same and aligned. In the up-down case, the interaction potential is normalized to the absolute value of its extremum, and in the side-by-side case to $U_0 = q_1 q_2 / 4 |a|$ .}
    \label{fig:Potential}
\end{figure}

The interaction potential in aligned case \eqref{eq:EMF_UD_int_Lagrangian2-2-1} has an extremum at $d=2 |a|$, representing an equilibrium state. The interaction potential in this case is depicted in Figure \ref{fig:PotentialPlus} and \ref{fig:PotentialMinus} for charges having the same or opposite signs, respectively.

For two magic fields in a side-by-side configuration, the interaction potential is obtained from \eqref{eq:EMF_UD_int_Lagrangian2} as
\begin{align} \label{eq:EMF_UD_int_potential2}
 U_{\text{\tiny SS}} \ & = \  
\left(q_{{}_+}q_{{}_-} \right) \frac{d\,(d+2f)-(a_{{}_+}-a_{{}_-})^2}{2\,d\,f\,\sqrt{(d+2 f)^2 - 4 \,e^2}}
\end{align}
For long-range interactions $a_{{}_\pm}<<d$, the leading terms are
    \begin{equation}
    U_{\text{\tiny SS}} \approx \left(q_{{}_+}q_{{}_-} \right) \left[\,\frac{1}{d}+ \frac{(a_{{}_+}+a_{{}_-})^2}{2\, d^3}+ \mathcal O\left(\frac{1}{d^5}\right) \right]
    \, .
    \end{equation}
    In particular, we recover the Coulomb interaction in spin-zero limit $a_{{}_\pm} \to 0$ . 

As special cases, we consider the spins of interacting magic fields being either aligned equal $a_{{}_\pm} = a$ as well as anti-aligned equal $a_{{}_\pm} = \pm a$ in side-by-side configuration. For both cases, we have $f = f_0 :=\frac12\sqrt{d^2 - 4\, a^2}$ and $e=0$. In the aligned case, we obtain
\begin{align} \label{eq:EMF_UD_int_Lagrangian2-1}
U_{\text{\tiny SS}}\Big|_{a_{{}_\pm} = a} \ & = \ 
\left(q_{{}_+}q_{{}_-} \right)\frac{1}{2\,f_0} \ = \ \left(q_{{}_+}q_{{}_-} \right)\frac{1}{\sqrt{d^2 - 4\, a^2}} \ ,
\end{align}
whereas in the anti-aligned case, we have
\begin{align} \label{eq:EMF_UD_int_Lagrangian2-2a}
U_{\text{\tiny SS}}\Big|_{a_{{}_\pm} = \pm a} \ & = \ \left(q_{{}_+}q_{{}_-} \right)\frac{1}d \ .
\end{align}
The interaction potential in aligned case \eqref{eq:EMF_UD_int_Lagrangian2-1} is depicted in Figure \ref{fig:PotentialPlus_ss} and \ref{fig:PotentialMinus_ss} for charges having the same or opposite signs, respectively.

\section{Conclusion and discussion} \label{sec:Conclusion}

The paper is devoted to the analysis of the electromagnetic magic field, the $G\! \to\! 0$ limit of the Kerr-Newman solution of Einstein-Maxwell equations and its interactions. We have analyzed the analytic continuation of the complex potential for a single magic field, as well as the superposition of two such fields. 
Although the self-energy of the magic field diverges, its Lagrangian is finite. 
This feature, which is contrary to the Coulomb field, admits the derivation of the interaction potential from the total Lagrangian of the superposed fields.

The interaction potential of two magic fields
has the Coulomb potential as the leading order term for large distances with corrections due to the spin parameters appearing at order $1/d^3$. For small distances, there is a qualitative difference from the Coulomb case.
For some configurations, we find, in contrast to the Coulomb interaction potential, a potential well with a diameter comparable to the spin $a$ of the field.

The work in this paper raises several questions that will be addressed in future work. As has been discussed above, the nature of the Kerr-Newman solution and the magic field exhibit several interesting analogies with the electron. The lowest electromagnetic multipole moments of the magic field and those of the electron are closely similar, which raises questions about the higher multipole moments of the electron, which yet have to be measured experimentally, as well as the role of the self-gravitating nature of the electron in determining its multipole structure. Here, one may draw an interesting parallel to uniqueness results for stationary electrovac black holes. 

The analogy with the electron is further emphasized by the fact that the analytic continuation of the magic field exhibits spin-$1/2$ features, 
demonstrated by the $4\pi$ angle needed to go around the singularity, cf. Figure \ref{fig:4pi_curve} and the discussion around it. 
The scale of the potential well found in the interaction potential could lead to an analogy of electron-positron pairs as well as orbiting configurations. These issues will be further discussed in a separate paper, where also the motion of the system of two magic fields will be considered.

Finally, it is worth mentioning the fact that the magic field is related to infinitesimal Kerr by the double copy procedure. There is a close analogy between linearized gravity and electromagnetism \cite{Barnett_2014,aghapour2021helicity}. From this point of view, it is interesting to consider the interaction potential for infinitesimal Kerr.  Some aspects of gravitational spin interaction can be found in\cite{wald1972gravitational,mashhoon2000gravitational}. It is also interesting to study the implications of such interactions for the post-Minkowskian black hole scattering \cite{Damour2016, Vines_2018, Vines_etal_2019}.

\subsection*{Acknowledgement}
KR thanks Lars Samuelsson and Mikael von Strauss for their valuable contributions at the early stages of this project.
SA \& LA thank the Institut Henri Poincaré (IHP), where part of this work was carried out during the program Quantum and Classical Fields Interacting with Geometry. SA is supported by the Humboldt Research Fellowship.

\appendix
\section{Interaction Lagrangian using the separating plane method}\label{sec:app_SP_metod}

In this appendix, we use another approach to analyze the interaction Lagrangian, which shows more clearly that the singular sets make no contribution to the interaction Lagrangian via the Laplacian terms in \eqref{eq:int_Lagrangian_potentials}. In this approach, we divide the space by a separating plane into two parts, each containing one of the singular sets, e.g. the point singularity of the Coulomb field or the ring singularity of the magic field. Applying Stokes' theorem then transforms the volume integral to surface integrals on closed surfaces around the singularities and on the separating plane. The interaction Lagrangian \eqref{eq:int_Lagrangian} then reads
\begin{align} \label{eq:int_Lagrangian_potentials_SP}
\mathcal L_{\text{int}} \ & = \ \frac{1}{\Omega} \int_{\mathcal V} \boldsymbol\nabla \mathcal Z_1 \cdot \boldsymbol\nabla \mathcal Z_2 \, \ud V \nonumber \\
\ & = \ \frac{1}{\Omega} \left( \int_{\mathcal V_1} \left[ \boldsymbol\nabla \cdot ( \mathcal Z_1 \, \boldsymbol\nabla \mathcal Z_2 ) - \mathcal Z_1 \, \boldsymbol\nabla^2 \mathcal Z_2  \right] \ud V
+\int_{\mathcal V_2} \left[ \boldsymbol\nabla \cdot ( \mathcal Z_2 \, \boldsymbol\nabla \mathcal Z_1)   - \mathcal Z_2 \, \boldsymbol\nabla^2 \mathcal Z_1 \right] \ud V \right)
\nonumber \\
& = \ \frac{1}{\Omega} \left(  \oint_{ \partial \mathcal V_1}  \mathcal{Z}_1 \boldsymbol \nabla \mathcal{Z}_2 \cdot \mathrm{d} \boldsymbol S
+ \oint_{ \partial \mathcal V_2} \mathcal{Z}_2 \boldsymbol \nabla \mathcal{Z}_1 \cdot \mathrm{d} \boldsymbol S \right)
\end{align}
where $\mathcal V= \mathcal V_1 \cup \mathcal V_2$ and $\boldsymbol\nabla^2 \mathcal Z_2 = 0$ in the whole region $\mathcal V_1$ and $\boldsymbol\nabla^2 \mathcal Z_1 = 0$ in the whole region $\mathcal V_2$.

For two Coulomb fields \eqref{eq:Coulomb_field}, we put the separating plane at $z=0$ so that $\mathcal V_1$ and $\mathcal V_2$ are the upper and lower halves of space, respectively. Then, each of the two parts of the interaction Lagrangian in \eqref{eq:int_Lagrangian_potentials_SP} has two sectors, the inner one around the singularity and the outer one containing $z=0$ plane and half of the spatial infinity. After taking the limit to shrink the surfaces at $z=\pm z_0$, the integration over the inner boundaries and the sectors at infinity of outer boundaries vanish. So, the only contribution to the integral is from the separating plane at $z=0$ from both sides. Therefore, the interaction Lagrangian for these two Coulomb fields is
\begin{align} \label{eq:Coulomb_int_Lagrangian_SP}
\mathcal L_{\text{int}} \ & = 
\ \frac{1}{4 \pi} \left(
-\int_{\mathcal B_+} %
\mathcal Z_+\,\frac{\partial \mathcal Z_-}{\partial z} \,  d S
+ 
\int_{\mathcal B_-} %
\mathcal Z_-\,\frac{\partial \mathcal Z_+}{\partial z} \,  d S \right)
\nonumber \\
& =  \ \left(q_{{}_+} q_{{}_-}\right)\times 2 \int_0^\infty \frac {\chi\, d\chi}{(\chi^2 + z_0^2)^2} = \left(q_{{}_+} q_{{}_-}\right)\frac{1}{d} \ ,
\end{align}
where $\mathcal B_\pm$ are upper and lower sides of the separating plane at $z=0$, and $d S = \chi d\chi d\phi$ is the area element in polar coordinates.

For two magic fields in the up-down configuration, when we adopt a similar method by 
locating the separating plane at $z=0$,
and considering the topological structure of the whole Riemann surface depicted in Figure \ref{fig:TopoInteractingMF}, which results in the overall factor $4\times 1/\Omega=1/4\pi$ and using the pillboxes like those in Figure \ref{image_up_down_2}, the interaction Lagrangian yields to
\begin{align} \label{eq:EMF_UD_int_Lagrangian_SP}
\mathcal L_{\text{int}} \ & = 
\ \ \frac{1}{4 \pi} \left(
- \int_{\mathcal D_+^+} %
\mathcal Z_+\,\frac{\partial \mathcal Z_-}{\partial z} \,   d S
+\int_{\mathcal D_+^-} %
\mathcal Z_+\,\frac{\partial \mathcal Z_-}{\partial z} \, d S
-
\int_{\mathcal B_+} %
\mathcal Z_+\,\frac{\partial \mathcal Z_-}{\partial z} \,  d S
\nonumber \right.
\\
& \qquad\qquad\quad \left.
-\int_{\mathcal D_-^+} %
\mathcal Z_-\,\frac{\partial \mathcal Z_+}{\partial z} \,   d S
+\int_{\mathcal D_-^-} %
\mathcal Z_-\,\frac{\partial \mathcal Z_+}{\partial z} \, d S
+
\int_{\mathcal B_-} %
\mathcal Z_-\,\frac{\partial \mathcal Z_+}{\partial z} \,  d S \right)
\, ,
\end{align}  
with contributions
\begin{align}
    \int_{\mathcal D_+^+} 
    \mathcal Z_+\,\frac{\partial \mathcal Z_-}{\partial z} \,   d S &= \lim\limits_{z\to z_0^{+}} \int_{0}^{|a_{{}_+}|} 
    \frac{-2 \pi(z+z_0-i a_{{}_-}) \chi d \chi}{\sqrt{\chi^2+(z-z_0-ia_{{}_+})^2}\left(\chi^2 +(z+ z_0- i a_{{}_-})^2 \right)^{3 / 2} } \nonumber \\
    &= \lim\limits_{\epsilon\to 0^{+}} \frac{-2\pi(2z_0-i a_{{}_-})}{(-ia_{{}_+})^2-(2z_0-ia_{{}_-})^2}\frac{\sqrt{(\epsilon-ia_{{}_+})^2}}{\sqrt{(2z_0-ia_{{}_-})^2}} \nonumber \\
    &= -\frac{2\pi i a_{{}_+}}{a_{{}_+}^2+(2z_0-ia_{{}_-})^2} \ .
\end{align}
\begin{align}
    \int_{\mathcal D_+^-} 
    \mathcal Z_+\,\frac{\partial \mathcal Z_-}{\partial z} \,   d S &= \lim\limits_{z\to z_0^{-}} \int_{0}^{|a_{{}_+}|} 
    \frac{-2 \pi(z+z_0-i a_{{}_-}) \chi d \chi}{\sqrt{\chi^2+(z-z_0-ia_{{}_+})^2}\left(\chi^2 +(z+ z_0- i a_{{}_-})^2 \right)^{3 / 2} } \nonumber \\
    &= \lim\limits_{\epsilon\to 0^{-}} \frac{-2\pi(2z_0-i a_{{}_-})}{(-ia_{{}_+})^2-(2z_0-ia_{{}_-})^2}\frac{\sqrt{(\epsilon-ia_{{}_+})^2}}{\sqrt{(2z_0-ia_{{}_-})^2}} \nonumber \\
    &= \frac{2\pi i a_{{}_+}}{a_{{}_+}^2+(2z_0-ia_{{}_-})^2} \ .
\end{align}
\begin{align}
    \int_{\mathcal B_{+}} 
    \mathcal Z_+\,\frac{\partial \mathcal Z_-}{\partial z} \,   d S &= \lim\limits_{z\to 0^{+}} \int_{0}^{+\infty} 
    \frac{-2 \pi(z+z_0-i a_{{}_-}) \chi d \chi}{\sqrt{\chi^2+(z-z_0-ia_{{}_+})^2}\left(\chi^2 +(z+ z_0- i a_{{}_-})^2 \right)^{3 / 2} } \nonumber \\
    &= \frac{2\pi(z_0-i a_{{}_-})}{(-z_0-ia_{{}_+})^2-(z_0-ia_{{}_-})^2}\left(1-\frac{\sqrt{(-z_0-ia_{{}_+})^2}}{\sqrt{(z_0-ia_{{}_-})^2}}\right)\nonumber \\
    &=\frac{2\pi}{-2z_0-ia_{{}_+}+ia_{{}_-}}\ .
\end{align}
\begin{align}
    \int_{\mathcal D_-^+}
    \mathcal Z_-\,\frac{\partial \mathcal Z_+}{\partial z} \,   d S &= \lim\limits_{z\to -z_0^{+}} \int_{0}^{|a_{{}_-}|} 
    \frac{-2 \pi(z-z_0-i a_{{}_+}) \chi d \chi}{\sqrt{\chi^2+(z+z_0-ia_{{}_-})^2}\left(\chi^2 +(z- z_0- i a_{{}_+})^2 \right)^{3 / 2} } \nonumber \\
    &= \lim\limits_{\epsilon\to 0^{+}} \frac{-2\pi(-2z_0-i a_{{}_+})}{(-ia_{{}_-})^2-(-2z_0-ia_{{}_+})^2}\frac{\sqrt{(\epsilon-ia_{{}_-})^2}}{\sqrt{(-2z_0-ia_{{}_+})^2}} \nonumber \\
    &= \frac{2\pi i a_{{}_-}}{a_{{}_-}^2+(-2z_0-ia_{{}_+})^2} \ .
\end{align}
\begin{align}
    \int_{\mathcal D_-^-} 
    \mathcal Z_+\,\frac{\partial \mathcal Z_-}{\partial z} \,   d S &= \lim\limits_{z\to -z_0^{-}} \int_{0}^{|a_{{}_-}|} 
    \frac{-2 \pi(z-z_0-i a_{{}_+}) \chi d \chi}{\sqrt{\chi^2+(z+z_0-ia_{{}_-})^2}\left(\chi^2 +(z- z_0- i a_{{}_+})^2 \right)^{3 / 2} } \nonumber \\
    &= \lim\limits_{\epsilon\to 0^{-}} \frac{-2\pi(-2z_0-i a_{{}_+})}{(-ia_{{}_-})^2-(-2z_0-ia_{{}_+})^2}\frac{\sqrt{(\epsilon-ia_{{}_-})^2}}{\sqrt{(-2z_0-ia_{{}_+})^2}} \nonumber \\
    &= -\frac{2\pi i a_{{}_-}}{a_{{}_-}^2+(-2z_0-ia_{{}_+})^2} \ .
\end{align}
\begin{align}
    \int_{\mathcal B_{-}} 
    \mathcal Z_-\,\frac{\partial \mathcal Z_+}{\partial z} \,   d S &= \lim\limits_{z\to 0^{-}} \int_{0}^{+\infty} 
    \frac{-2 \pi(z-z_0-i a_{{}_+}) \chi d \chi}{\sqrt{\chi^2+(z+z_0-ia_{{}_-})^2}\left(\chi^2 +(z- z_0- i a_{{}_+})^2 \right)^{3 / 2} } \nonumber \\
    &= \frac{2\pi(-z_0-i a_{{}_+})}{(z_0-ia_{{}_-})^2-(-z_0-ia_{{}_+})^2}\left(1-\frac{\sqrt{(z_0-ia_{{}_-})^2}}{\sqrt{(-z_0-ia_{{}_+})^2}}\right)\nonumber \\
    &=\frac{2\pi}{2z_0-ia_{{}_-}+ia_{{}_+}}\ .
\end{align}
where we have used the primitive integral
\begin{align}
\int \frac{\xi_{1} \chi d \chi }{\sqrt{\chi^2+\xi_{2}} \left(\chi^2+\xi_{3}\right)^{3/2}}   = - \frac{\xi_1 }{\xi_2-\xi_3} \frac{\sqrt{\chi^2+\xi_2}}{\sqrt{\chi^2+\xi_3}} + c \, ,
\end{align}
for $\xi_{1},\xi_{2},\xi_{3} \in \mathbb{C}$ and $\chi \in \mathbb{R}$ and skipped the charges. In these calculations, we have used a similar analysis to that in \eqref{eq:squre_root_anlyisis1}
Adding up all the above terms according to \eqref{eq:EMF_UD_int_Lagrangian_SP} and putting back the charges, the interaction Lagrangian is obtained as
\begin{align}
    \mathcal L_{int} \  &= \ (q_{{}_+}q_{{}_-})\,\frac1{2z_0+ia_{{}_+}-ia_{{}_-}}\left(\frac{ia_{{}_+}}{2z_0-i(a_{{}_+}+a_{{}_-})}+1-\frac{ia_{{}_-}}{2z_0+i(a_{{}_+}+a_{{}_-})}\right)\nonumber \\
    &= \ (q_{{}_+}q_{{}_-})\,\frac{d}{d^2+(a_{{}_+}+a_{{}_-})^2} \ ,
\end{align}
the same as \eqref{eq:EMF_UD_int_Lagrangian2a}. In this way, we showed that the Laplacian terms in \eqref{eq:int_Lagrangian_potentials} have no contribution. One can perform the same analysis for the magic fields in the side-by-side case and reach to the same conclusion.

\section{Coulomb interaction in bi-spherical coordinates}\label{sec:app_Coulomb_bispherical}
In this appendix, we use the bi-spherical coordinates to calculate the Coulomb interaction from Maxwell Lagrangian. The advantage of this coordinate system is that its singular points match that of the problem, i.e. the location of the two point charges. Consider two Coulomb fields with their singular points at $x=\pm x_0$. The adapted bi-spherical coordinates are defined as
\begin{align}
    x = \frac{x_0 \sinh \eta}{\cosh \eta - \cos \sigma}\ , \quad
    y = \frac{x_0 \sin \sigma \sin\psi}{\cosh \eta - \cos \sigma}\ , \quad
    z = \frac{x_0 \sin \sigma \cos \psi}{\cosh \eta - \cos \sigma}\ ,
\end{align}
where 
\begin{align}
    dx \,dy\, dz = \frac{x_0^3 \sin \sigma}{\cosh \eta - \cos \sigma}\,d\eta\,d\sigma\,d\psi \,
\end{align}
and the whole space $V=\mathbb{R}^3$ is covered by $\eta \in (-\infty,\infty)$, $\sigma\in [0,2\pi)$ and $\psi\in[0,\pi]$. 

The interaction Lagrangian of the two Coulomb fields yields
\begin{align}
    L_{\text{int}} & = \frac1{4\pi} \int_V \, E_1 \cdot E_2 \, \ud V \nonumber \\
     & = \frac{q_{{}_+}q_{{}_-}}{4\pi} \int_V \frac{(x^2+y^2+z^2 - x_0^2) \ud x \ud y \ud z}{\left((x-x_0)^2+y^2+z^2\right)^{3/2}\left((x+x_0)^2+y^2+z^2)\right)^{3/2}} 
     \nonumber\\
     & = \frac{q_{{}_+}q_{{}_-}}{4\pi} \int_V \frac{\sin\sigma\,\cos\sigma}{4x_0\,(\cos\eta - \cos\sigma)}\ud\eta\ud\sigma\ud\psi
     \nonumber\\
     & = (q_{{}_+}q_{{}_-})\, \frac1d \ ,
\end{align}
where $d=2x_0$ is, like before, the distance between the singular points.

\section{Magic Interaction in oblate spheroidal coordinates}
In this appendix, we calculate the interaction Lagrangian of up-down magic fields in Oblate spheroidal coordinates. We take the inner surfaces in \eqref{eq:EMF_int_Lagrangian} to be the infinitesimal oblate spheroids $\mathcal S_\pm$ around two singular rings so that
\begin{align} \label{eq:EMF_int_Lagrangian_SP}
\mathcal L_{\text{int}} \ & = \ \frac{1}{8\pi} \left(\oint_{\mathcal S_+} \boldsymbol \nabla(\mathcal{Z}_+\,  \mathcal{Z}_-) \cdot \mathrm{d} \boldsymbol S
+  \oint_{\mathcal S_-} \boldsymbol \nabla (\mathcal{Z}_+\,  \mathcal{Z}_-) \cdot \mathrm{d} \boldsymbol S \right) \ .
\end{align}
Using a pair of oblate spheroidal coordinates and considering the corresponding representation of magic fields \eqref{eq:PotentialOblate}, the integrands in \eqref{eq:EMF_int_Lagrangian_SP} are a gradient of the product
\begin{equation}\label{eq:magic_updown_OS}
\Ptl_+\, \Ptl_ - =\frac{q_{{}_+}\,q_{{}_-}}{(r_{{}_+}-i\, a_{{}_+} \cos \theta_{{}_+})\left(r_{{}_-} -i\, a_{{}_-} \cos \theta_{{}_-} \right)} \ . 
\end{equation}
To compute the first term in \eqref{eq:EMF_int_Lagrangian_SP}, we shall rewrite the above expression only in terms of "$+$" coordinates (i.e. the oblate spherical coordinates associated to the upper magic field), using the relations to the cylindrical coordinates as
\begin{subequations}
\begin{align}\label{eq:os_coordinates}
&\chi=\sqrt{r_{{}_+}^{2}+a_{{}_+}^{2}} \, \sin \theta{{}_+} = \sqrt{r_{{}_-}^{2}+a_{{}_-}^{2}} \, \sin \theta{{}_-} \, , \\
&z= r_{{}_+} \cos \theta_{{}_+} = r_{{}_-} \cos \theta_{{}_-} - d \, ,
\end{align} 
\end{subequations}
where $d$ is the distance between the centers of the singular rings in the up-down configuration.
The potential for the lower magic field is then a square-root function of a polynomial for $\xi = \cos\theta_+$ as
\begin{align}
\label{eq:ppolynom}
\left(r_{{}_-} -i\, a_{{}_-} \cos \theta_{{}_-}\right)^2 & = \chi^{2}+(z + d- i \,a_{{}_-})^{2} \nonumber\\
& = -a_{{}_+}^{2}\, \xi^{2}+2\,b(r_{{}_+})\, \xi+ c(r_{{}_+})=:p(r_{{}_+},\xi) \, ,
\end{align}
where $b(r):=\left(d-i\, a_{{}_-}\right)r$ and $c(r) = r^2+a_{{}_+}^{2}+\left( d-i\, a_{{}_-}\right)^{2}$. The polynomial \eqref{eq:ppolynom} has two roots
\begin{align}
    \xi_\pm(r) = \left( b(r) \pm \sqrt{ b^2(r) + a_{{}_+}^2 \, c(r)} \, \right) / \, a_{{}_+}^2 \ .
\end{align}

Using the polynomial \eqref{eq:ppolynom}, the product expression \eqref{eq:magic_updown_OS} takes the form
\begin{equation}
    \Ptl_+\, \Ptl_ - =\frac{q_{{}_+}\,q_{{}_-}}{(r_{{}_+}-i\, a_{{}_+} \cos \theta_{{}_+})\sqrt{p(r_{{}_+}, \cos\theta_{{}_+})}} \ .
\end{equation}
With the area element on the spheroid $\mathcal S_+$ to be $dS=(r_{{}_+}^2+a_{{}_+}^2 )\sin\theta\ud\theta\ud\phi$, we have
\begin{align}\label{eq:S+_int}
    \oint_{\mathcal S_+} \boldsymbol \nabla(\mathcal{Z}_+\,  \mathcal{Z}_-) \cdot \mathrm{d} \boldsymbol S =  \lim_{r_{{}_+}\!\to \,0 }\int_0^{2\pi}\int_0^\pi \frac\partial{\partial r_{{}_+}} \left(\frac{q_{{}_+}\,q_{{}_-}}{(r_{{}_+}-i\, a_{{}_+} \cos \theta_{{}_+})\sqrt{p(r_{{}_+}, \cos\theta_{{}_+})}} \right) (r_{{}_+}^2+a_{{}_+}^2 )\sin\theta\ud\theta\ud\phi \ .
\end{align}
We can commute the derivative $\partial/\partial r_{{}_+}$ with the the integrals. A primitive function for the resultant integral is
\begin{eqnarray}
\lefteqn{
    \int \frac{\mathrm{d} \xi}{(r-i\, a\, \xi) \sqrt{-a^{2}\left(\xi-\xi_{+}\right)\left(\xi-\xi_{-}\right)}} =
}&&
\nonumber
\\
&&
\frac{2 \sqrt{\xi-\xi_{+}} \sqrt{\xi-\xi_{-}}}{\sqrt{ r-i\,a \,\xi_{+}}\sqrt{r-i\,a \,\xi_{-}} \sqrt{-a^{2}\left(\xi-\xi_{+}\right)\left(\xi-\xi_{-}\right)}  }
\arctanh \left\{\frac{ \sqrt{r-i\,a \,\xi_{+}}\sqrt{\xi-\xi_{-}}}{ \sqrt{r - i\,a\, \xi_{-}}\sqrt{\xi-\xi_{+}}}\right\} + c 
\ .
\end{eqnarray}
Analysing for the principal branch, the integral \eqref{eq:S+_int} leads to
\begin{align}
    \oint_{\mathcal S_+} \boldsymbol \nabla(\mathcal{Z}_+\,  \mathcal{Z}_-) \cdot \mathrm{d} \boldsymbol S =
    \frac{4\pi\,q_{{}_+}q_{{}_-}}{d - i \, (a_{{}_+} + a_{{}_-})} \ .
\end{align}
A similar computation for the second term in \eqref{eq:EMF_int_Lagrangian_SP}, leads to
\begin{align}
     \oint_{\mathcal S_-} \boldsymbol \nabla(\mathcal{Z}_+\,  \mathcal{Z}_-) \cdot \mathrm{d} \boldsymbol S 
     = \frac{4\pi\,q_{{}_+}q_{{}_-}}{d + i \, (a_{{}_+} + a_{{}_-})} \ .
\end{align}

Finally, the total interaction Lagrangian \eqref{eq:EMF_int_Lagrangian} for two magic fields in up-down positions is obtained as
\begin{align} \label{eq:EMF_UD_int_Lagrangian2a_SP}
\mathcal L_{\text{int}} \ & = \  \frac{1}{8\pi} \left( \frac{4\pi\,q_{{}_+}q_{{}_-}}{d - i \, (a_{{}_+} + a_{{}_-})} + \frac{4\pi\,q_{{}_+}q_{{}_-}}{d + i \, (a_{{}_+} + a_{{}_-})} \right)
\nonumber\\
\ & = \ (q_{{}_+}q_{{}_-})\,\frac{d}{d^2 + (a_{{}_+} + a_{{}_-})^2} \ , 
\end{align}
as we have found in \eqref{eq:EMF_UD_int_Lagrangian2a} in cylindrical coordinates.

\bibliographystyle{unsrt} 
\bibliography{references}

\end{document}